\documentclass[12pt,preprint]{aastex}

\newcommand{\lessim}{\mathrel{\hbox{\rlap{\hbox{\lower4pt\hbox{$\sim$}}}\hbox{$<$}}}}
\newcommand{\gtsim}{\mathrel{\hbox{\rlap{\hbox{\lower4pt\hbox{$\sim$}}}\hbox{$>$}}}}

\newcommand{\logg}{$\log{g}$}
\newcommand{\loggs}{$\log{g}$ }
\newcommand{\teff}{$T_{\rm eff}$}
\newcommand{\teffs}{$T_{\rm eff}$ }
\newcommand{\ebv}{${\rm E(B-V)}$}

\shorttitle{White dwarf initial-final mass relation}
\shortauthors{Salaris et al.}

\begin{document}

\title{Semi-empirical white dwarf initial-final mass relationships: \\
    a thorough analysis of systematic uncertainties \\
    due to stellar evolution models}

\author{Maurizio Salaris\altaffilmark{1}}
\affil{Astrophysics Research Institute, Liverpool John Moores
University,
Twelve Quays House, Egerton Wharf, Birkenhead CH41 1LD, UK}
\email{ms@astro.livjm.ac.uk}

\and

\author{Aldo Serenelli\altaffilmark{1}}
\affil{Institute for Advanced Study, Einstein Drive, Princeton,
           NJ~08540, USA}

\and
\author{Achim Weiss}
\affil{Max-Planck-Institut f\"ur Astrophysik,
  Karl-Schwarzschild-Str.~1, 85748~Garching, Germany}

\and

\author{Marcelo Miller Bertolami\altaffilmark{2}}
\affil{Facultad de Ciencias Astron\'omicas y Geof\'isicas,
Universidad Nacional de La Plata, Paseo del Bosque S/N, (1900) La
Plata, Argentina}

\altaffiltext{1}{Max-Planck-Institut f\"ur Astrophysik, Karl-Schwarzschild-Str.~1, 85748~Garching, Germany}
\altaffiltext{2}{IALP-CONICET, La Plata, Argentina}

\begin{abstract}
Using the most recent results about white dwarfs in 10 open clusters,
we revisit semi-empirical estimates of the initial-final mass relation in star clusters,
with emphasis on the use of stellar evolution models.
We discuss the influence of these models on
each step of the derivation. One intention of our work is to use
consistent sets of calculations both for the isochrones and the white
dwarf cooling tracks. The second one is to derive the range of systematic
errors arising from stellar evolution theory. This is achieved by
using different sources for the stellar models and by varying physical
assumptions and input data. We find that systematic errors, including
the determination of the cluster age, are
dominating the initial mass values, while observational uncertainties
influence the final mass primarily. After having determined the systematic
errors, the initial-final mass relation allows us finally to draw
conclusions about the physics of the stellar models, in particular
about convective overshooting.
\end{abstract}



\section{Introduction}

The initial-final mass relation (IFMR) for low- and intermediate-mass stars
is an important input for many astrophysical problems. Given the initial Main
Sequence (MS) mass of a formed star, the IFMR
provides the expected mass during its final White Dwarf (WD) cooling stage,
and is an estimate of the total mass lost by the
star during its evolutionary history. A correct assessment of the IFMR
is very important when predicting, for example, the
chemical evolution history of stellar populations, or
their mass-to-light ratio (defined as the ratio of the mass of
evolving stars plus remnants -- WDs, neutron stars and black
holes -- to the integrated luminosity of the population), and in
general for any problem related to the origin and evolution of gas in stellar
populations. It is also a crucial item when using the WD luminosity
functions to determine the age of stellar populations
\citep[see, e.g.,][]{prad07}. It also provides an
empirical test concerning the upper initial mass value $M_{\rm up}$ of
stars developing degenerate carbon-oxygen cores, and thus a lower
limit of the mass range of core-collapse supernovae. For a thorough
review of the history of the IFMR, we recommend \citet{weide00}.

Theoretical estimates of the IFMR are still prone to large
uncertainties. This is due to our poor knowledge of the
efficiency of mass loss processes for low- and intermediate-mass
stars but also to uncertainties in the predicted size
of CO cores during the Asymptotic
Giant Branch (AGB) evolution, resulting mainly from uncertainties in the
treatment of the thermal pulse phase, with the associated third
dredge-up, hot bottom burning \citep[see, e.g.,][]{iberen83} and also
the treatment of rotation \citep{domi96}. In fact, the recognition
that intermediate-mass stars may lose the largest part of their mass
during the AGB phase stems from early IFMRs
\citep{koeweid80}.

Starting with the pioneering work by
\cite{weid77}, (semi-)empirical routes have been followed to establish
the IFMR independently of theoretical modelling the AGB phase,
leading to a series of global determinations
of the IFMR by, e.g., \citet{weidkoe83}, \citet{weid87},
\citet{weide00}, \citet{fer05}, and many others.

Methods to estimate the IFMR that make use of the smallest number of
assumptions are probably those based on the
use of WDs harbored in star clusters
\citep[but see also][for a very recent study of the IFMR based on
WDs in common proper motion pairs]{cata07}.
Thanks to a large amount of observational effort by various authors
\cite[see, e.g.][]{kr93, kr96, clav01, dob04, kal05,
bars05, dobbie06} the recent study of the IFMR  by \citet{fer05}
makes use of 40 DA (hydrogen-atmosphere)
WDs belonging to 7 open clusters.
Even more recently \citet{kalir08} have added a few more data
points below the low-mass end
(initial masses below $\sim$2.0 $M_{\odot}$)
of the \citet{fer05} sample, by including WDs
detected in the old open clusters NGC6819, NGC7789 and
NGC6791, while \citet{rubin08} have provided WD data for
the $\sim$200~Myr old cluster NGC1039.
Therefore, the observational material is increasingly
covering not only more objects, but also a larger range of cluster --
and thus WD -- ages and metallicities, such that even differential
investigations may become feasible in the near future. At the same
time, high-resolution, high signal-to-noise spectra and improved
spectral analysis methods reduce the intrinsic errors in the
determination of the WD masses. It is therefore time to investigate
the systematic errors arising from stellar evolution theory and
models that enter these semi-empirical methods.
This is the main object of this work.

IFMR determinations based on cluster WDs work as follows:
after detection, spectroscopic estimates of the WD
surface gravity $g$ and \teff\ are needed. For a fixed $g$-\teff\
pair, interpolation within a
grid of theoretical WD models covering a range of masses provides the
mass $M_{\rm f}$ and the age $t_{\rm cool}$ of the WD.
Independent theoretical isochrone fits to the Turn Off (TO)
luminosity in the cluster Color-Magnitude-Diagram (CMD)
provide an estimate of the cluster age
$t_{\rm clus}$.  The difference  $t_{\rm clus}-t_{\rm cool}$  is equal  to the
lifetime
$t_{\rm prog}$ of the WD progenitor, from the MS until the start of the WD
cooling. Making use of mass-lifetime relationships from theoretical
stellar evolution models, the initial progenitor mass $M_{\rm i}$ is
immediately obtained from $t_{\rm prog}$.

It is clear how and where theoretical models play
an important role in this procedure. WD cooling models
plus progenitor evolutionary tracks and isochrones
are needed to determine the various $t_{\rm clus}$, $t_{\rm cool}$
and $t_{\rm prog}$. Given the very short timescale of the AGB evolution
(see also next section for a short discussion on this issue) AGB
modelling and its related uncertainties do not
play a role in these types of analyses, but
uncertainites entering the previous evolutionary phases and the WD
cooling stage can affect the derived IFMR.
Another important point is the issue of consistency. First of all,
the isochrones
employed to determine $t_{\rm clus}$ should be computed
starting from the same evolutionary tracks adopted to
determine $M_{\rm i}$ from the estimated value of $t_{\rm prog}$.
Consistency is not usually achieved in existing estimates of the
IFMR. Even in the most recent studies of the IFMR, in several cases
cluster ages are obtained from earlier studies that have employed models
different from the ones used to estimate $M_{\rm i}$.
Secondly, all cluster ages should be determined following the same
method, and -- most importantly -- employing the same set of
isochrones. Instead, results from different sources for the various
clusters are used.

As an example, we summarize briefly the work by \citet{fer05}.  The
Hyades age has been taken from \citet{perr98}, who had calculated
their own isochrones with an updated version of the CESAM code
\citep{morel97}.  In case of M35 the age quoted in Table~1 of
\citet{hippel05} was employed; this age was derived summarizing the
results of three different previous works by other authors. The same
is true for the Pleiades, where the assumed cluster age comes from a
synthesis of the results of several groups, obtained with different
generations of stellar evolution models.
The isochrones by \citet{schall92} led \citet{meynet93} to an age of
141~Myrs for NGC2516, which was quoted by \citet{fer05} as
158~Myr. The remaining clusters in that sample also have ages from
various isochrone sources, which span more than a decade of progress
in stellar evolution modelling, most notably the significant improvements
in opacities and the equation of state. The cooling tracks came from
\citet{font01}, and the $t_{\rm prog}$-- $M_{\rm i}$ relation from
\citet{gir02}, taken from the tracks with the metallicity closest
to the adopted one, but without interpolation between compositions. It
is obvious that there is a certain degree of arbitrariness in the use
of stellar evolution theory.
This inherent inconsistency is probably the result
of the fact that most works on the IFMR are more focused on
acquiring new and improved data with an emphasis on the
internal errors of their derived \teff\ and $M_{\rm f}$.

The aim of the present work is instead to concentrate on the stellar
evolution models needed for deriving the IFMR. It is obvious from the previous discussion
that a consistent use of stellar models and procedures about how to
employ them is required, to achieve a
higher accuracy. We therefore will make use of calculations of stellar
models, that cover all phases from the MS to the WD
cooling phase. They will be used for the appropriate metallicity of
each cluster, in order to have at hand accurate and homogeneous
isochrones and progenitor lifetimes. This will provide an
IFMR with the highest internal consistency achieved so far.

To get a measure of the errors introduced by using different codes and
input  physics, we  also repeat  the procedure  by using  alternative  sets of
models for the WD cooling and/or progenitor and cluster ages.
While this gives a first indication about the size of the
systematic errors resulting form different stellar model sources, we
finally quantify the influence of various physical effects important
for both the isochrone and WD cooling calculations. Using Monte Carlo
methods we determine the total extent of the systematic
errors. This allows us also to verify which features  of the
IFMR are robust and which ones depend on present model uncertainties.

Using the robust results, we can finally exclude some of the stellar models
because they lead to internal contradictions. One of our conclusions is that
models without overshooting from convective cores on the MS do not
yield self-consistent results. On the other hand, the parametrization of
overshooting during the AGB
phase, which influences the growth of the carbon-oxygen core, cannot be the
same in all convective layers as it is at the convective core during
the MS phase, because it produces final masses inconsistent with
observations. We thus can also reduce
uncertainties in the physics of stellar interiors.

The structure of the paper is as follows: In Sect.~2 we outline the
procedure to obtain an IFMR, summarize the observational input data
we have used for the WDs and the parent clusters, and
present the set of the stellar models used for determining the
reference IFMR as well as for the evaluation of the systematics.
Section~3 then concentrates on the estimates of the global error
bars, while Sect.~4 contains the results and a comparison with
theoretical IFMRs. A summary and conclusions follow in Sect.~5.

\section{IFMR determination \label{sec:ifmr}}

As discussed in the Introduction, the method we follow to determine
the IFMR makes use of WDs in star clusters. Observations have to
provide us with:
\begin{itemize}
\item{Surface gravity $g$ and effective temperature \teff\ of
   individual WDs.}
\item{Observed CMDs, estimates of [Fe/H] and \ebv\ for the clusters
   harboring the observed WDs.}
\end{itemize}
Theory has to provide:

\begin{itemize}
\item{Sets of stellar evolutionary tracks,  plus isochrones --
   transformed to the CMD --  for a range of age and
   [Fe/H] values, to determine -- in conjuction with theoretical, empirical
   or semi-empirical distance determination methods --  the cluster ages, plus
   stellar lifetimes.}
\item{WD cooling tracks covering a range of masses, to
   provide the cooling ages of the observed WDs.}
\end{itemize}

Concerning the cooling tracks, the metallicity of
the WD models is irrelevant as the envelopes are of pure H (He and
metals have settled below the envelope due to diffusion) and the
previous core composition may influence only the chemical profile
of the core. This will be investigated in a broader context later
in this study.

The method works as already summarized in the Introduction;
for each cluster WD with an estimated pair of $g$ and \teff\
values,
linear interpolation within a grid of theoretical WD models provides
the mass $M_{\rm f}$ (mainly fixed by the value of $g$)
and the  age $t_{\rm cool}$ (mostly  determined by the value  of \teff) of
the WD.
Isochrone fits to the TO region of the observed ${\rm V-(B-V)}$ CMD
determine the cluster age $t_{\rm clus}$.
If the actual cluster [Fe/H] does not correspond to
any of the values in the isochrone grid used, we
determined the age with the two closest grid values bracketing the
cluster [Fe/H]. The final cluster age is then obtained by linear
interpolation between these two values.

The difference $t_{\rm clus}-t_{\rm cool}$ represents the lifetime
$t_{\rm prog}$ of the WD progenitor, from the MS until the start of the WD
cooling. Theoretical mass-lifetime relationships for the cluster
metallicity, obtained from the same evolutionary calculations, finally
provide the initial progenitor mass $M_{\rm i}$. We
employ a quadratic interpolation
in mass and a linear interpolation in [Fe/H] among the available model
grids to determine $M_{\rm i}$ from the value of $t_{\rm prog}$.

A potential inconsistency in this procedure is the effect of the
duration of the AGB thermal pulse phase on the estimate of $M_{\rm i}$.
Theoretical models without inclusion of the thermal pulse phase, or
predicting an IFMR different from the one obtained semi-empirically,
might be providing inconsistent thermal pulse lifetimes,
hence wrong values of the total
lifetimes, that in turn affect the derivation of $M_{\rm i}$. Luckily, this
effect is negligible; estimates of thermal pulse lifetimes
obtained from synthetic AGB modelling \citep{wg98, mg07} show that the
duration of this phase is typically at most within a few Myr (decreasing
sharply with increasing total mass) even in extreme
cases  of the  largest possible  growth of  the CO  core. In  fact,  from full
evolutionary stellar models, \cite{Serfuk07}
have shown  that the duration of  the AGB phase (including  both the early-AGB
and the thermally  pulsating phases), for stars with  initial masses between 1
and 8~M$_\odot$, is at most 1\% of the MS lifetime.

\subsection{WD data}

Our reference data set comprises 52 WDs in 10 open clusters, plus
Sirius~B, for a total of 53 DA objects, as reported in
Table~\ref{WDsample}.  We considered the same cluster sample as in the
recent work by \citet{fer05}, 
that contains data for WDs in the Pleiades \citep{fer05},  
Hyades and Praesepe \citep{clav01, dob04}, 
NGC2516 \citep{kr96}, NGC3532 \citep{kr93}, NGC2099 \citep{kal05},  
NGC2168 \citep{wbk04, fer05}.  We wish to notice that our Pleiades sample 
contains only the object listed by \citet{fer05}. We did not 
include data for additional two WDs (GD~50 and PG~0136+251) discussed by 
\citet{doetal06}, because their membership of the Pleiades 
is not absolutely certain.

To this cluster sample we added 
NGC6819, NGC7789,
and NGC1039. For the WDs in the first two clusters, \teff\ and $g$ values
are taken from \citet{kalir08}, and from \citet{rubin08} for the last
one.

The \teff, $g$ and corresponding error estimates for the clusters in
common with \citet{fer05} are the same as in their Table~1, with a few
exceptions. Values of \teff\ and $g$ for two Praesepe WDs (0837+185
and 0837+218), identified as WD \#13 and \#14 in Table~\ref{WDsample},
have been updated following \citet{dobbie06}. The same authors provide
data for four additional WDs in the same cluster (ID numbers from 16
to 19 in Table~\ref{WDsample}), that we have included in our
analysis. And finally, we also use their data for the single Pleiades
WD.
 
As an additional note about the cluster WD sample, 
our referee (K. Williams) has pointed out that the Hyades 
WD~0437+138, that appears as WD \#8 in Table~\ref{WDsample}, has been incorrectly included as 
a DA object in \citep{fer05} study. In fact, it is a DBA WD, with 
log($n_H/n_{He}$)=$-$4.5 in the atmosphere. To assess the errors in the derived 
mass and cooling age for this object when using DA models, 
we have made the following test. We have compared log($g$) values  
derived from DB and DA models at the observed \teff\ of WD \#8, 
for the two extreme cases of 0.55 and 1.0~$M_{\odot}$ (the mass 
we obtain from the analysis with DA models is $\sim$0.77$M_{\odot}$). 
Our adopted DB models have been computed as described in \citet{scg01}. 

For both masses the difference 
in log($g$) derived from the DA and DB tracks is smaller than the 1$\sigma$ error bar 
on the observed value for WD \#8 (as listed in Table~\ref{WDsample}). 
Also the difference in cooling ages at the observed 
\teff\ is well below the error bars we derive from our analysis, as reported in Table~\ref{WDage1}.
This implies that the analysis of WD \#8 with DB models would provide mass and cooling age compatible 
(within the 1$\sigma$ errors) with the results obtained 
from the DA models employed in this analysis. The reason is that the mass-radius relation is not strongly 
affected by the atmospheric composition, and  also at the relatively high \teff\ of 
this object the evolutionary timescales of DA and DB models are similar. 
Moreover, given that this WD does not have a pure-He envelope, 
differences will be even more negligible.

In addition to a large sample of cluster WDs, we also include
Sirius~B in our analysis. The data for Sirius~B \citep[also
included in][]{fer05} in Table~\ref{WDsample} are from the
recent determination by \citet{bars05}.
From the estimated
mass and radius of the primary MS star (Sirius~A) one can derive its age
by comparison with stellar evolutionary tracks in the mass-radius plane.
We assumed [Fe/H]=0.0 for Sirius~A, following
\citet{lieb05}, to which we associated an 0.10~dex 1$\sigma$ error.
The age of the star is obtained from
models bracketing the metallicity of the system (if different from the values
provided by the employed isochrone grid). Linear interpolation in [Fe/H]
between the two bracketing results gives then the age for the appropriate [Fe/H].
The age of Sirus~A provides the age of the system
$t_{\rm sys}$. Sirius~B cooling time $t_{\rm cool}$ and mass are determined
from the spectroscopic $T_{\rm eff}$  and $g$ estimates, as described before. The difference
$t_{\rm sys}$-$t_{\rm cool}$ provides the age of the Sirius~B progenitor,
hence $M_{\rm i}$ after using a grid of theoretical lifetimes for solar
metallicity.

\subsection{Cluster data}

The sources for the cluster [Fe/H], E(B-V), and their CMDs used for age
determinations, are listed in Table~\ref{clsource}.
 
Values for [Fe/H] have been taken mainly from the compilation by 
\citet{grat00}, who presents a collection of  
[Fe/H] estimates for a large sample of open clusters, 
recalibrated against results from high-resolution spectroscopy.
Alternative estimates have been adopted for NGC2099, which is not 
in the \citet{grat00} compilation, 
NGC6819 and NGC1039, for which recent high-resolution spectroscopic 
measurements have appeared in the literature (see Table~\ref{clsource}).

Reddenings are taken from the compilation by \citet{lot01}. These are weighted averages of 
estimates obtained with various methods, all based essentially 
on color-color relationships. For NGC6819 we employed a very recent E(B-V) determination
based on spectroscopy \citep{brag01}, and 
for the Pleiades reddenings to individual stars have been adopted, as discussed in 
\citet{per03a}. The case of NGC7789 is discussed below.

In case of Hyades and Praesepe we have assumed
-- as usual -- zero reddening. Throughout this paper we use
${\rm{A_V}}=3.2~{\rm E(B-V)}$.
We adopted a 0.02~mag uncertainty in \ebv\ when no error is quoted by
the source of the data.

To estimate the age of the whole cluster sample we have
followed a two-step procedure
where we tried to minimize the inputs from theoretical isochrones.
First, cluster distance moduli have been determined using a fully
empirical MS-fitting to the
${\rm V-(B-V)}$ CMDs, that makes use of the sample of field dwarf with accurate
$Hipparcos$ parallaxes and [Fe/H] presented by \citet{per03a}. 
Details about the method can be found in the same paper. 

Here we just recall that for each cluster a MS fiducial line 
has been derived, by plotting color histograms in  
V-magnitude bins, and considering the mode of the resulting distribution. In this way 
one minimizes the impact of unresolved binaries. The resultant points are fitted 
to a cubic function that provides the cluster fiducial line.
After  applying (empirical)  color  corrections that  account  for the  actual
cluster metallicity 
and for the reddening E(B-V), the field dwarf template sequence is shifted 
in magnitude to match the MS fiducial line, and the 
apparent distance modulus is obtained by $\chi^2$ minimization.
For objects like  NGC2168 or the Pleiades, with an  extremely well defined and
thin (in color) MS, 
the separation of the binary sequence is very evident, and a cubic fit directly to the 
single star MS is performed in order to derive the fiducial line.

In case  of NGC3532 we could not  find available data with  good photometry in
the MS magnitude 
range of our template field dwarf sample. 
The  distance modulus of  this cluster  has been  obtained by  comparison with
Praesepe, that has 
the same [Fe/H] within errors (and also a comparable age).
The larger  (as compared  to our  other estimates) error  bar on  its distance
(hence on its age) reflects 
this less direct determination of the distance modulus.

As a final remark, we wish to notice that, as discussed i.e.\ in \citet{per03a}, a 
simultaneous MS-fitting to both VI and BV data can provide both 
the  cluster distance modulus  and  reddening. This  procedure has  been
applied (using the same 
field dwarf sample, that has homogeneous BVI photometry) to NGC7789, employing additionally 
the VI data by \citet{gim98}. Our adopted E(B-V) for NGC7789, used also in \citet{per03b},  
has been determined  in this way. This method to determine the reddening 
could not  however be applied to the whole cluster sample because 
of the lack of suitable BV and VI photometry for all clusters.

For a few clusters we have used distances already published in the
literature, obtained with the same technique and the same $Hipparcos$
field dwarf sample. More in detail, for the Hyades we have used the result by
\citet{per03a}, that is in
perfect agreement with the $Hipparcos$ distance by \citet{perr98}.
For the Pleiades we have used the distance modulus determined by
\citet{per05}, while for NGC7789 we have employed the result
by \citet{per03b}. The cluster  NGC6791, which is also part of the
investigation by \citet{kalir08} was not considered in our analysis
because its metallicity [Fe/H]$\sim$+0.45 \citep{gra06}
is higher than the upper [Fe/H] range of  the
field dwarf sample ([Fe/H]=+0.30), and therefore we could
not apply this empirical method to determine its distance. Since we
did not want to introduce a second, inconsistent method, we
disregarded this cluster.

Second, isochrone fits to the TO region (in the ${\rm V-(B-V)}$ plane),
employing the distance moduli derived in the previous step, determine
the cluster age.

\begin{figure}
\epsscale{1.00}
\plotone{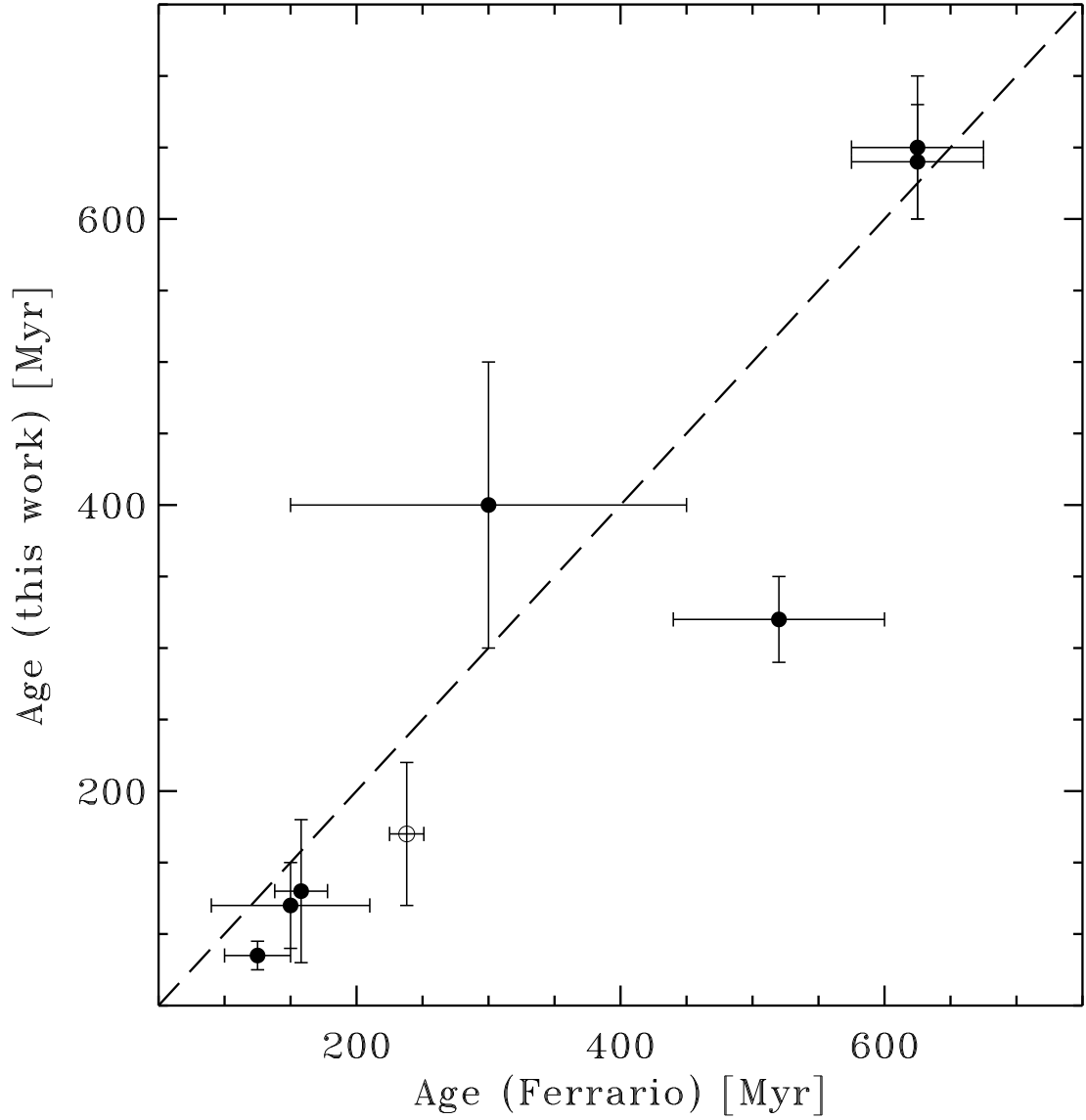}
\caption{Comparison of the cluster ages determined with
the BaSTI overshooting isochrones, with the ages adopted from
the literature by  \citet{fer05}, for the clusters in  common. The dashed line
displays the 1:1 
relation, and the errors are those given by \citet{fer05} or as listed
from Table~\ref{Ages}. The open circle shows the results for Sirius~A.
\label{clustagecomp}}
\end{figure}

The final ages for various choices of the isochrone grid, as well as
the adopted cluster [Fe/H] and \ebv\ values are reported in
Table~\ref{Ages}. Age uncertainties are due to both distance (which is in turn
affected by the assumed [Fe/H] and \ebv\ values and their errors)
and [Fe/H] uncertainties.
Figure~\ref{clustagecomp} compares the ages
obtained with the BaSTI overshooting isochrones (that throughout the
paper will be considered as our reference set of models to determine
the IFMR; see Sect.~2.3) with the ages adopted by \citet{fer05} for
the clusters in
common (plus Sirius~A).  In spite of the dishomogeneity in terms of
methods, data and isochrones employed, the two sets of ages compare
well, within the respective error bars. The only exception occurs for
NGC2099 (M37), for which our derived distance modulus is 0.4
magnitudes larger than implicitly assumed in \citet{fer05} --
who adopt the age estimate by \citet{kal01a} --  leading to
a smaller cluster age (320~Myr as compared to 520~Myr). However, if we
adopt a lower metallicity and reddening as suggested by
\citet{kal05} and discussed in Sect.~2.4 (see Table~\ref{Ages}) we
derive an age of 550~Myr for NGC2099, in close agreement with
\citet{fer05}. Obviously, at this time the uncertainties in the
cluster parameters dominate the determination over those from using
different theoretical isochrones, which constitutes our first result.

\subsection{Models}

Our reference IFMR is defined by the use of the following set of
models. We refer the reader to the original papers quoted below for
details of the models and calculations.

-- \citet{pietr04} models from the MS to the AGB
with the inclusion of core overshooting during the central H-burning phase
(hereafter BaSTI models with overshooting), to
estimate in a consistent way both $t_{\rm clus}$ and $t_{\rm prog}$.

-- \citet{sala00} WD cooling models (hereafter S00 WD models) to
estimate $t_{\rm cool}$. The code employed to compute these WD models
is the same as for the progenitors.  Some major sources of input
physics in the WD models (see S00) are different from what has been
employed in BaSTI models, most notably equation of state and
low-temperature opacities. This is somewhat unavoidable, given the
different range covered by many physical parameters throughout WD
structures, as compared to their progenitors. It is very important to
remark here about the internal CO stratification of the WD models. The
CO profile in the inner core of a WD is built up during the central
He-burning phase, whereas the more external part of the WD core is
processed during the early AGB phase and the thermal pulses.  Current
uncertainties in the treatment of mixing in stellar interiors and in
some key reaction rate (i.e., the ${\rm C}^{12}+\alpha$ reaction)
introduce additional uncertainties in the predicted CO profiles.  S00
models employ a CO stratification obtained from evolutionary models of
solar metallicity progenitors, from the MS until the
completion of the first thermal pulse \citep{sal97}. The core mass and
associated CO profile at the end of the pulse were taken as
representative of the final WD object. When estimating the total
error budget in the IFMR, we have accounted for the uncertainties in
the CO profiles by using WD models with different C/O mass fraction
ratios in their cores. This will be described in detail in
Sect.~\ref{sec:uncert}. We anticipate, however, that these
uncertainties are very small since all the WD stars in our sample have
not entered yet the crystallization phase, where the detailed core
composition could have a larger effect.

We have studied systematic effects on the IFMR relation due to:

i) Different model grids. We redetermine the
IFMR employing \citet{gir00} models from the MS to the AGB (hereafter
Padua models; they also include core overshooting but with a different
formalism  compared to BaSTI), keeping the WD models unchanged (S00).

ii) Effect of overshooting. We redetermine the
IFMR employing \citet{pietr04} models from the MS to the AGB without
core overshooting (hereafter BaSTI no-OV models), again keeping the
WD models unchanged (S00).

iii) Different WD model grids. We redetermine the IFMR
employing the WD models computed with the LPCODE (hereafter LPCODE models;
details of the LPCODE are given in \citealt{alt03}),
keeping the progenitor models unchanged (BaSTI with overshooting).
This tests not only different codes,
but also -- in an unsystematic way -- variations of the physics
employed and of the WD chemical stratification.

iv) Systematic variation in WD input physics and chemical profiles. We
redetermine the IFMR employing the LPCODE WD models, by varying the CO-core
stratification, H-envelope thickness, electron conduction opacities and
neutrino energy loss rates.  In this case, the progenitor models are
unchanged throughout (BaSTI with overshooting).

\subsection{The reference IFMR}

We present in this section our reference IFMR, while we defer to
subsequent sections the discussion of the uncertainties involved in
the IFMR determination.  We use the S00 WD models to derive the WD (final)
masses and cooling ages.  Using our reference cluster ages
(fifth column in Table~\ref{Ages}) and the BaSTI models with
overshooting we construct the IFMR by deriving the
progenitor (initial) masses. Figure~\ref{fig:ref_ifmr} displays this reference
semi-empirical IFMR. Error bars include all
uncertainty  sources we  consider. We  defer a  detailed discussion  to
Sections~\ref{sec:uncert}~and~\ref{sec:results}, but note here that the most
relevant sources of uncertainty are due to the following effects:
(i) errors in the WD \teff\ and $g$ values; (ii) uncertainties in
the cluster [Fe/H]; these affect the age through their influence on
the derived cluster distance (mainly) and on the isochrones to be used for the
age determination; (iii) errors in the reddening \ebv, which
influence the age determination through the effect on the cluster
distance.  In case of one star
(WD \#38 in NGC2099) the progenitor age results to
be negative. We did not consider this object in the analysis of the IFMR.

We have performed at this stage a simple check of our reference IFMR,
comparing the distribution of $M_{\rm i}$ and $M_{\rm f}$ values with
\citet{fer05} -- disregrading the attached error bars -- by means of a
bi-dimensional Kolmogorov-Smirnov (KS) test \citep{ptvf92}. We
considered only clusters (plus Sirius) in common between the two
studies. As customary, we accept the existence of a significant
difference between the two IFMRs if the probability $P$ that the two
samples of ($M_{\rm i}$, $M_{\rm f}$) pairs were drawn from different
distributions was higher than 95\%.  We obtain $P$ values smaller than
50\%, highlighting the statistical agreement between our reference
IFMR and \citet{fer05} results for the clusters (plus Sirius~B) in common.

We have also performed analytical fits to the data. In doing this, we have left out
the data from NGC2099 because our results for this cluster show some anomalies,
e.g.  a progenitor with negative age (WD \#38 mentioned above), a
0.45~M$_\odot$ WD with a progenitor mass above 8~M$_\odot$ (WD \#37) and some
other issues to be discussed later in the paper.

In Figure~\ref{fig:ref_ifmr} we show two  analytic fits to the data. The first
one, linear, gives
\begin{equation}
M_{\rm f}= 0.084 M_{\rm i} + 0.466
\end{equation}
and a reduced  chi-squared $\chi^2_r=3.7$. Approximate  values for the 1$\sigma$ rms
dispersion  about  this   relation  are:  0.075~M$_\odot$  below  2~M$_\odot$,
0.12~M$_\odot$  between  2.7 and  4~M$_\odot$,  0.09~M$_\odot$  between 4  and
6~M$_\odot$  and then  it increases  almost linearly  up to  0.20~M$_\odot$ at
around 8.5~M$_\odot$.

Guided by theoretical models, which
show  a  break   in  the  slope  of  the  IFMR   at  around  4~M$_\odot$  (see
Fig.~\ref{mimfbastisoo}), we have also performed a piecewise linear fit with a
pivot point at $M_{\rm i}=4~{\rm M_\odot}$ that yields
\begin{equation}
M_{\rm f} = \left\{ \begin{array}{lllc} 0.134 M_{\rm i} + 0.331 & & & 1.7~
{\rm M}_\odot \leq M_{\rm i} \leq 4~{\rm
M}_\odot\\
0.047 M_{\rm i} + 0.679 & & & 4~{\rm M}_\odot \leq M_{\rm i}.
\end{array} \right.
\end{equation}
This fit gives $\chi^2_r=2.7$ at the expense of introducing only one more
free parameter (we force the fit to be continuous at the pivot point).  In
this case, the dispersion is reduced to 0.05~M$_\odot$ below 2~M$_\odot$
and to 0.09~M$_\odot$  above  6~M$_\odot$,   while  it  remains  basically
unchanged anywhere else. The reader should keep in mind that these cluster data
do not constrain the IFMR at masses lower than about 1.7 M$_\odot$.
As a consequence, the present analytic relations, particularly the piecewise
linear fit, should not be extrapolated towards lower initial masses.

We could attempt more complicated fits to the data, but in our opinion
the uncertainties in
mass determinations, and possibly the existence of an intrinsic spread in
the IFMR to  be discussed below, render such efforts probably unnecessary. In fact, we
have found  polynomial fits  beyond second order  do not produce  any relevant
improvement in the goodness of fit. 
 
As an additional check, we performed a 
direct  comparison of  our  fits  with  the polynomial  relationship  from
\citet{fer05}, that was derived 
by combining an assumed star formation rate and initial mass function 
for the Galactic Disk, with the observed field WD mass distribution. 
For masses between 2.5  and 6.5~M$_\odot$ -- the range of initial
masses covered in their study -- at a given $M_{\rm i}$, differences in the
$M_{\rm f}$ values are less than
0.04~M$_\odot$, well below the dispersions given above.

Another point that we wish to address is the presence of a possible
intrinsic spread in the reference IFMR, as noticed by \citet{reid96} in an
analysis of Praesepe WD population.  With intrinsic spread we mean a signature
of differential mass loss among stars with the same mass and same initial
chemical composition. \citet{cata08} have investigated very recently the spread
around their mean semi-empirical IFMR (determined from cluster WDs and
WDs in common proper motion pairs) and found no clear cut
indication that is either due to a spread in mass loss or to some
other  individual  stellar properties  (e.g.  the  thickness  of the  envelope
layers) that vary between objects.

To investigate this issue with our data, we made the
following test.  We considered objects in a single cluster, namely
Praesepe, with $M_{\rm i}$ values
in agreement within their respective 1$\sigma$ error bars.  There are
8 stars with $M_{\rm i}$ between 3.045 and 3.543~$M_{\odot}$ satisfying this
constraint, and we assume that -- within the present error bars --
they share the same $M_{\rm i}$, equal to the average value $\langle M_{\rm i}
\rangle=3.207~M_{\odot}$.
For these objects, the average WD mass is $\langle M_{\rm f} \rangle
=0.769~M_{\odot}$ with 1$\sigma$ rms equal to 0.045$M_{\odot}$. To
test whether this observed dispersion in $M_{\rm f}$ is due exclusively to
the errors in the $M_{\rm f}$ estimates, we built a synthetic sample by
drawing randomly 10000 $M_{\rm f}$ values for each of the 8 objects,
according to a Gaussian distribution  centred on $\langle M_{\rm f}
\rangle=0.769~M_{\odot}$, with a 1$\sigma$ spread given by their
individual error bars.  The synthetic distribution has obviously
$\langle M_{\rm f} \rangle=0.769~M_{\odot}$, but a 1$\sigma$ rms equal only
to 0.028$M_{\odot}$. We compared statistically the $\sigma$ values of
the two distributions by means of an F-test, that returned a probability
of $P$=96.3\% that $\sigma$ of the synthetic sample is different from
(smaller than) the observed one.  By fixing, as customary, a threshold
of $P$=95\% to accept the existence of a real difference between the
two $\sigma$ values, our simple test seems to point to the existence
of a spread in the IFMR, at least in this cluster.  Adding 5 more
objects in the Hyades and in NGC3532 (the [Fe/H] estimates
for these two clusters plus Praesepe span a range of only $\sim$0.10~dex,
and  overlap  within  their  associated   $\sim$ 2$\sigma$  error  bars,  see
Table~\ref{Ages})
with initial masses within the same
overlapping $M_{\rm i}$ range and repeating the previous analysis does not
change this conclusion. In fact, for this enlarged sample, $\langle
M_{\rm f} \rangle =0.729~M_{\odot}$, with a 1$\sigma$ rms equal to
0.10$M_{\odot}$.  The probability that the corresponding synthetic
sample -- constructed in the same way as for the 8 Hyades WDs -- has a
different $\sigma$ than observed is $P$=96.9\%.

However, this evidence for an intrinsic spread (i.e.\ due to a
significantly varying mass loss at a fixed mass and chemical composition)
in the IFMR is weakened if we take into account theoretical
expectations for near-solar metallicities, displayed in
Fig.~\ref{mimfbastisoo}. Although the $M_{\rm i}$ values
estimated for these 13 WDs (or just the 8 Praesepe objects discussed above)
overlap within the 1$\sigma$ error bars, the same error bars suggest
that they may also cover a range of values. Just
considering the range covered by the central $M_{\rm i}$ estimates for these
13 WDs, theoretical IFMRs \citep[see also another completely
independent theoretical IFMR for solar metallicity in Fig.~3 of][]{prad07}
generally display a steep rise of $M_{\rm f}$ with
$M_{\rm i}$.  Over the $M_{\rm i}$ range between 3.05 and 3.5~$M_{\odot}$, the
theoretical IFMRs of Fig.~\ref{mimfbastisoo} predict an increase in
$M_{\rm f}$ by about $0.1\,M_\odot$, which corresponds to $\sim 2\sigma$ of
the observed $M_{\rm f}$ spread in the 8 Praesepe objects, and to $\sim
1\sigma$ of the observed spread in the 13-object enlarged
sample. Without accounting for errors in the estimate of $M_{\rm f}$, the
predicted change within the $M_{\rm i}$ range allowed by the error bars of
the estimated initial masses can potentially explain the observed
$M_{\rm f}$ spread. This would imply that the observed spread is not the
result of star-to-star mass loss variations, but due to,
as predicted from theory, the steep increase of $M_{\rm f}$ with $M_{\rm i}$
in the relevant mass range.

Moving this analysis to different $M_{\rm i}$ ranges can hardly provide
additional constraints on the existence of an intrinsic spread in the IFMR,
because of either smaller numbers of objects
in comparable $M_{\rm i}$ intervals (when moving to lower $M_{\rm i}$ values) or a
combination of much larger error bars on $M_{\rm i}$ and small numbers (when
moving to higher $M_{\rm i}$ values). It may however be interesting to
notice the situation at $M_{\rm i} \sim 1.8~M_{\odot}$, where we have 5
objects (belonging to NGC7789 and NGC6819) with formally the same
$M_{\rm i}$, small errors in both $M_{\rm i}$ and $M_{\rm f}$ and theoretical
predictions for an IFMR essentially flat (see
Fig.~\ref{mimfbastisoo}).  By applying the same statistical analysis
described before we do not find any statistically significant spread
in the IFMR for this sample.

Before concluding this section we discuss briefly another test of our
reference IFMR.  In the case of NGC2099 we have considered the alternative
value [Fe/H]=$-$0.20 and reduced reddening (E(B-V)=0.23) used by
\citet{kal05} and given by \citet{del02}. Note however that Deliyannis
(private communication) considers this as a possible but as yet
unconfirmed value for [Fe/H].  Using this alternative metallicity and
reddening we derive an age for NGC2099 of 550~Myr which, compared to
320~Myr derived with [Fe/H]=0.09, has a dramatic effect in the
inferred initial masses (Fig.~\ref{fig:m37}).  Results for NGC2099 stars
(\#27-\#38) are repeated in the last part of Table~\ref{MiMftab} but
in this case corresponding to the low [Fe/H]. As expected from the
relatively large change in the inferred cluster age, changes in $M_{\rm i}$
due to the different cluster parameters generally exceed the
uncertainties in $M_{\rm i}$ determination. 
 
This test, based on estimates that are admittedly not confirmed, 
is just to exemplify in a quantitative way 
how crucial the assessment of the cluster [Fe/H] and E(B-V) values is. 

We will return to this issue in Sect.~4.

\section{Global error-bar estimates \label{sec:uncert}}

The method used to determine the IFMR has been described in
Section~\ref{sec:ifmr}, where we also presented our reference
IFMR. We now focus on a detailed determination of the
global uncertainties both in the initial and final masses for each
star in our sample. Since the global errors have contributions from
various sources, they have been estimated by means of Monte Carlo simulations.
In the following we detail the procedure followed for each star in our sample.

\subsection{White Dwarf Uncertainties \label{sec:mcwd}}

We have tried to cover all major sources of uncertainties in the
WD mass  and cooling age determinations, that are going to be
discussed in this section.
While we have used S00 models as our standard set of WD models, all
the WD evolutionary sequences to test the effect of WD input physics and
chemical stratification have been computed with the
LPCODE described below. This should not invalidate our procedure,
since we are using LPCODE models
only to determine differential changes in the WD properties (e.g. core
composition), not absolute values.
In what follows, $X$ represents the logarithmic values of either the WD
mass or the cooling age.

--  Observational  uncertainties.  The  central values  and
  $1\sigma$   uncertainties   for   \loggs   and   \teffs   are   listed   in
  Table~\ref{WDsample}. We leave them unmodified as they are given in
  the original papers quoted in Sect.~2.1.

-- Systematic  uncertainties  from  WD  cooling  tracks. In  addition  to  our
  standard choice of WD
  cooling tracks (S00), we have specifically computed a completely independent
  set of WD cooling models with the LPCODE. The aim is to
  estimate the  {\it systematic} uncertainties  in the inferred  WD properties
  that arises from using different WD models.  We define the  $1\sigma$
uncertainty by
\begin{equation}
\sigma_{\rm syst}(X)=|X_{\rm S00}-X_{\rm LP}|;
\label{eq:wdsys}
\end{equation}
where $X_{\rm S00}$ is derived from S00 models and $X_{\rm LP}$ from
models computed with  the LPCODE. WD models  computed with the LPCODE
have  been used  in  \cite{Serfuk07}, where  the  evolution of  a set  of
stellar  models  from   1  to  8~M$_\odot$  and  solar   metallicity  has  been
consistently  followed from  the main  sequence phase  up to  the  WD
phase, including the thermal pulse (TP) AGB phase. It should be noted that, in addition
to differences in numerics, $\sigma_{\rm syst}(X)$ also accounts for
differences in input physics between the two sets of models. While S00
and LPCODE models have used the same neutrino energy loss rates
\citep{itoh96}, conductive opacities in the CO cores \citep{itoh93} and very
similar hydrogen envelope thickness and core composition, they differ
in other  relevant aspects such as  the equation of state  and low temperature
opacities
(see \citealt{alt03} and \citealt{sala00} for details about the LPCODE and S00
input physics respectively) and the inclusion of gravitational
settling in the model envelopes (S00 models mimic the effects of
gravitational settling by adopting a pure hydrogen envelope and fixed
chemical H/He and He/C/O interfaces while in LPCODE the chemical
profiles are evolved according to time-dependent equations of element
diffusion).

--  Neutrino energy loss rates.  For the WD evolutionary sequences we
are considering in this work, plasma neutrinos always contribute at least 90\%
of the  total neutrino emission. Uncertainties in  the theoretical calculation
of plasma neutrino emission rates,  within the framework of the standard model 
of particle physics, are at most  a few percent (Fukugita, private communication) and
thus we do not expect them  to represent an important source of uncertainty in
determining WD cooling ages. 
Several additional cooling
mechanisms   that  could  operate   in  WD   interiors  have   been  proposed,
e.g.  enhanced  neutrino cooling  due to a  non-null neutrino  magnetic dipole
moment, axion production or other particles like 
Weakly Interacting Massive Particles (see \citealp{raf_book}  
for a complete discussion of these mechanisms). It
is not our aim here to consider the effects on the WD properties --  
particularly on the cooling rate -- of all such mechanisms, especially because
they remain purely hypothetical\footnote{Among the
  non-standard   cooling   mechanisms,    axions   are   probably   the   best
  motivated. Recently,  \cite{bish08} and \cite{ise08} have  used WD pulsation
  properties 
  and the luminosity function of local Disk WDs, respectively, to constrain the 
  axion mass $m_a$.  While \cite{bish08} find an upper limit
  $m_a=13$~\mbox{meV}, \cite{ise08}  rule out values as  low as 10~\mbox{meV}
  and give a preferred value of $m_a=5$~\mbox{meV}. With this mass value, axion
  cooling is never the dominating WD cooling mechanism (contrary to neutrino
  cooling, that  dominates for hot  WDs) but it  is strong enough that  it can
  have a moderate effect on WD age determinations for WDs with luminosities in
  the range $-1.2 < \log L/{\rm L_\odot} \lesssim -1.7$.}. We do
not want, however, ignore the 
possibility that some additional cooling might be present. Consequently, 
we have adopted a more conservative point of view and
have computed additional WD cooling models where standard neutrino energy loss
rates  \citep{haft94,itoh96} have  been multiplied  by  factors of  2 and  0.5
respectively to take into account the possibility of such additional
cooling mechanisms.
These large factors were 
chosen to represent our conservative approach.
We define the WD mass (and cooling age) $1\sigma$ uncertainties due to
neutrino energy losses  as
\begin{equation}
\sigma_\nu(X)=\frac{|X_{\rm 2}-X_{\rm 0.5}|}{2},
\label{eq:wdneu}
\end{equation}
where the subindices refer
to quantities corresponding  to the WD  cooling models with  the modified
neutrino emission rates as described above.

-- Conductive  opacity.  For the  physical  conditions  and  composition of  a
representative set 
   of WD models, we have
   computed conductive opacities according to both Itoh and collaborators
   (\cite{itoh93} and  references therein -- our references  set of conductive
   opacities) 
   and Potekhin and collaborators (see \citealt{cass07} for the most up to
   date calculations). We find differences between both sets of
   calculations to be usually below 20\% for the range of temperatures
   and densities relevant to the WD models used in this paper.  In
   analogy with the neutrino emission rates, we have estimated the effect of
   electron conduction uncertainties by computing
   additional  WD  models  where   the  conductive  opacities  from  Itoh  and
   collaborators 
   have been multiplied by factors of 1.25 and 0.8 and
   have used these models to define the WD mass and cooling age
   $1\sigma$ uncertainties due to errors in conductive opacities
   calculations as
\begin{equation}
\sigma_\kappa(X)=\frac{|X_{\rm 1.25}-X_{\rm 0.8}|}{2}.
\label{eq:wdkap}
\end{equation}

-- WD core composition. The mass fraction of the central oxygen
   $X_{\rm O}$ in S00  WD models ranges from 0.65 up to  0.80 depending on the
   mass
   of the WD.
   The internal CO stratification is however subject to uncertainties
   due to the uncertainty on the ${\rm C}^{12}+\alpha$ reaction rate
  (see, e.g. S00) and, very important,
  the treatment of mixing  (semiconvection, breathing pulses and the mechanism
   to
  suppress them) during the central He-burning phase \citep{stra03}.
  These effects are larger than the  expected change in CO profiles due to the
  small metallicity
  range covered by the cluster sample.
  In our analysis we have been conservative and have defined the
  $1\sigma$ uncertainty due uncertainties in the core composition as
\begin{equation}
\sigma_{\rm core}(X)=\frac{|X_{\rm 0.9}-X_{\rm 0.3}|}{2},
\label{eq:wdcor}
\end{equation}
   where $X_{\rm 0.9}$ and $X_{\rm 0.3}$ are derived from WD cooling
   tracks where the original chemical profiles have been rescaled to have
   $X_{\rm O}$= 0.9 and 0.3 at the center, respectively.

-- Hydrogen   envelope   thickness  ($M_{\rm   H}$).   As   shown  by,   e.g.,
   \citet{prov98}
   the H-envelope thickness of DA WDs displays a range of values.
   To include this effect in our error analysis, in addition to the standard
   thick envelopes considered in our LPCODE calculations
   (usually $10^{-4}$~M$_\odot$, but larger values
   were used   for WD masses below 0.6~M$_\odot$) we have computed also
   cooling sequences with thinner H-layers of $10^{-6}$~M$_\odot$.
   Similar  to  the  above  procedures,  we  have  defined  the  corresponding
   $1\sigma$ errors
\begin{equation}
\sigma_{\rm env}(X)=\frac{|X_{\rm thick}-X_{\rm thin}|}{2};
\label{eq:wdenv}
\end{equation}
where $X_{\rm thick}$ and $X_{\rm thin}$ are derived from WD
models with thick and thin hydrogen envelopes respectively.
We have performed
an additional test to explore very thin hydrogen envelopes (as in
\citealt{cata08}) by  computing a  0.6  and a  1~M$_\odot$  WD models  with
hydrogen envelopes  of $M_{\rm H} \sim  10^{-10}$~M$_\odot$. When compared
to models with thick envelopes, we have found 
that differences in the cooling ages never exceeded 11\% in both cases 
and are usually at the level of 7\% for the 0.60~M$_\odot$ model and
5\% for the more massive one in the range of effective temperatures of
interest to this work. Asteroseismology studies of ZZ Ceti stars 
disclose a wide range of
hydrogen envelope thickness, ranging between $\sim 10^{-4}$~M$_\odot$
and $\sim 10^{-9}$~M$_\odot$ \citep[see, e.g.,][]{ck08}.
In this regard,
the changes  in the cooling ages due to uncertainties in the  H-envelope
thickness quoted above can probably be taken as robust upper limits.

For  neutrino  emission  rates,  conductive opacities,  core  composition  and
hydrogen-envelope  thickness  we have  assumed  that uncertainties  distribute
normally. In  the case of  systematic uncertainties from different  WD cooling
tracks we have assumed uniform distributions of uncertainties
($f_X$) which are defined, in each case,
by $\int_{-\sigma_{\rm syst}}^{+\sigma_{\rm syst}}{f_X dX}=0.683$.  We
tabulate $\sigma_\nu(X)$, $\sigma_\kappa(X)$, $\sigma_{\rm core}(X)$,
$\sigma_{\rm env}(X)$ and $\sigma_{\rm syst}(X)$ in (\logg,\teff)
grids for using them in the Monte Carlo simulations. We summarize the
results for the WD uncertainties, both for mass and cooling age, in
Table~\ref{tab:uncert}, where for a 
small  subset  of (\logg,  \teff)  values,  we  give the  $1-\sigma$  fractional
uncertainties in WD mass and cooling age as defined
by  Eqs.~\ref{eq:wdsys}-\ref{eq:wdenv},  but transformed  to  linear scale  to
facilitate interpretation of the results.

\subsection{Progenitor Star Uncertainties \label{sec:mcms}}

Progenitor  masses $M_{\rm  i}$  are obtained  from  the estimated  progenitor
lifetime 
and stellar  models. In addition to  the uncertainty sources  discussed in the
previous section for $t_{\rm cool}$, the determination of $M_{\rm i}$ is affected by:

-- Cluster age. Progenitor lifetimes are obtained simply as
$t_{\rm prog}=t_{\rm clus} - t_{\rm cool}$, so uncertainties in $t_{\rm clus}$
directly affect the determination of $M_{\rm i}$. Central values and adopted 
$1\sigma$ uncertainties  for the cluster  ages determined for this  paper are
listed in
Table~\ref{Ages}.    As  our   standard  choice,   we  have   adopted  uniform
distributions $f_t(t_{\rm clus})$ for
cluster  age uncertainties,  where $f_t$  is a constant
defined by the simple relation $\int_{-\sigma_{t {\rm clus}}}^{+\sigma_{t {\rm
 clus}}} f_t \,dt_{\rm clus}=0.683$.

-- Stellar metallicity  (${\rm  \Delta[Fe/H]}$).  ${\rm  \Delta[Fe/H]}$
  affects the determination of
 $M_{\rm i}$ in two different ways.  The first and most important relates to the
 determination of the cluster age (mainly through the
 change in the distance modulus obtained via MS-fitting)
 and  has already been taken into account in
  the uncertainty of $t_{\rm clus}$.  Additionally, ${\rm \Delta[Fe/H]}$
 introduces uncertainty  in the determination of the  progenitor mass because,
  for different metallicities, a fixed $t_{\rm prog}$ corresponds to different
  progenitor masses. We  treat this effect by adopting  the central values and
  uncertainties in  the metallicities  listed in Table~\ref{Ages}  to generate
  stellar  metallicity distributions  (normal distributions  are  our standard
  choice).  For  a  fixed  $t_{\rm  prog}$  we  then  compute  $M_{\rm i}$  for  the
  corresponding metallicity distribution.

-- Different  stellar models  and isochrones. In  order to  estimate the
  systematic uncertainties arising from using different sets of stellar models
  and isochrones we have proceeded in a  similar fashion as in the case of the
  WD cooling tracks and define as the $1\sigma$ uncertainty
\begin{equation}
\sigma_{\rm syst}(M_{\rm i})=|M_{\rm i, BaSTI}-M_{\rm i, Padua}|;
\end{equation}
where  $M_{\rm i, BaSTI}$  represents the progenitor masses  derived with
  Basti stellar  models and  isochrones and $M_{\rm  i, Padua}$  those derived
  with Padua  models and isochrones.   We have adopted a  uniform distribution
  for systematic uncertainties. It should be noted here again that consistency
  in the calculations requires the same set of models and 
isochrones be used to derive $M_{\rm  i}$ and $t_{\rm prog}$, i.e. the cluster
age.
It is in principle not correct to adopt a cluster age derived with, let us say
the BaSTI isochrones, and then the  Padua stellar models to derive $M_{\rm i}$
from 
the calculated $t_{\rm prog}$.  As already stated in the Introduction, such
inconsistencies  are ubiquitous in  the current  literature and  can introduce
noticeable  systematic changes in  the derived  $M_{\rm i}$,  particularly for
$M_{\rm i} \gtsim 5$M$_\odot$.

We close this section by stressing that both codes, BaSTI
and Padua, are employing up-to-date physics input, and therefore are, at
least nominally, quite similar in their treatment of stellar evolution.
Systematic uncertainties
arising from code-to-code differences are therefore possibly
underestimated in the above error analysis, if they arise from
different physical assumptions, as, for example, the inclusion
or neglect of convective overshooting, which we did not include
in the discussion of this section (we provide separate results for models
without core overshooting).

When considering  models without core  overshooting, the systematic  effect of
using different models and isochrones was taken into  account with a slightly
different  approach because  Padua  calculations are  only  available for  one 
chemical composition, i.e. $Z$=0.019 and $Y$=0.273. 
This pair of ($Y,Z$) values corresponds 
to [Fe/H]=0.04 when considering the \citet{gn93} solar mixture 
(that provides $Z/X$=0.0245$\pm$0.005 for the present Sun) 
employed in those stellar model calculations. Also BaSTI models 
employ the \citet{gn93} solar mixture, and the ($Y,Z$) pair closest to the 
Padua one around the solar metallicity is $Z$=0.0198, $Y=$0.2734, 
that corresponds to [Fe/H]=0.06. 
In this  case, and  just for  estimating this
systematic uncertainty,  we have computed all 
progenitor masses for both BaSTI  and Padua non-overshooting models 
assuming a fixed  [Fe/H]=0.04 for Padua and [Fe/H]=0.06 for BaSTI. 
As  above, the  difference between  the masses has  been adopted  as a
measure of the systematic uncertainty and we have assumed this difference does
not  depend  on  the  actual  [Fe/H]  value. This  is  in  fact  a  reasonable
approximation and,  in addition, systematic uncertainties  from isochrones and
models turn out to be a  small contribution to the overall uncertainty budget,
so our simplified treatment should not affect our results and conclusions.

\subsection{Monte Carlo simulations}

For each star in our sample, the Monte Carlo (MC) simulations
consist of the following steps:

-- generate $ \{\log{g}_{\rm j},T_{\rm eff,j}\}_{\rm {j=1}}^{\rm N}$
distributions according to observational uncertainties and use S00 WD
models to get the initial WD mass $ \{M^0_{\rm f,j} \}_{\rm
  {j=1}}^{\rm N}$ and cooling age $ \{t^0_{\rm cool,j} \}_{\rm
  {j=1}}^{\rm N}$ distributions.  Here $N(=10^5)$ is the total number
of trials in the MC simulations and quantities with the $j$ suffix
denote results of each one of the trials;

-- For each trial $j$ include additional WD mass and cooling age
  uncertainties according to the details given in
  Section~\ref{sec:mcwd}. Each mass in the the final WD mass
  distribution $ \{M_{\rm f,j} \}_{\rm {j=1}}^{\rm N}$ (analogously
  for the WD cooling age) is given by
  \begin{equation}
    M_{\rm f,j}= M^0_{\rm f,j} + \sum_{\rm k=1,5} \delta M_{\rm j,k},
  \end{equation}
  where each $\delta M_{\rm j,k}$, the individual contributions to the
  WD mass global uncertainty, is randomly drawn from distributions
  constructed as described in Section~\ref{sec:mcwd}. The final WD
  cooling age distribution $\{t_{\rm cool,j} \}_{\rm {j=1}}^{\rm N}$
  is computed in the same way;

-- generate the cluster age distribution $\{t_{\rm clus,j} \}_{\rm
  {j=1}}^{\rm N}$ and then the progenitor age distribution $\{t_{\rm
  prog,j} \}_{\rm {j=1}}^{\rm N}$, where for each trial $t_{\rm
  prog,j} = t_{\rm clus,j} - t_{\rm cool,j}$;

-- From $\{t_{\rm prog,j} \}_{\rm {j=1}}^{\rm N}$ get the initial
  progenitor mass distribution $ \{M^0_{\rm i,j} \}_{\rm {j=1}}^{\rm
  N}$ using the BaSTI stellar models;

-- The final progenitor mass distribution $ \{M_{\rm i,j} \}_{\rm
  {j=1}}^{\rm N}$ is obtained by adding the star metallicity and
  stellar models systematic uncertainties, i.e. $M_{\rm i,j}= M^0_{\rm
  i,j} + \delta M_{\rm i,[Fe/H]} + \delta M_{\rm i,syst}$ as discussed
  in Section~\ref{sec:mcms}.

\section{Results and comparisons with theory \label{sec:results}}

\subsection{Global error assessment}

Table~\ref{WDage1} displays the results about cooling age and mass for all
the WDs in our sample. The central values have been
obtained using S00 WD models and the uncertainties are the result of
the MC simulations described in the previous section and thus include
{\sl all sources of uncertainty considered in this work}.  We also provide,
as additional information, the bolometric luminosity of the WDs, as
obtained from the interpolation among the WD model grid in \loggs and
\teff. It is important to notice that none of the objects have yet
entered the crystallization phase. This minimizes the effect of
uncertainties in the interior CO profiles, that play a very important
role during the phase separation associated to the crystallization
process (S00).

Our reference IFMR relation (BaSTI with core overshooting plus S00
models) is displayed in Table~\ref{MiMftab} (labelled ``ov'').  For
the WD masses, the
quoted uncertainties are the average of $\sigma_-(M_{\rm f})$ and
$\sigma_+(M_{\rm f})$ given in Table~\ref{WDage1}.  For the progenitor
masses we give the results of our Monte Carlo simulations, that show
strongly asymmetric error bars arising from the highly non-linear
relation between evolutionary lifetime and mass for MS
stars.  Results for BaSTI models without core overshooting are also shown
(labelled ``non-ov'').

As already discussed, our estimates of the global uncertainties in
the IFMR include systematic effects due to the use of different sets of stellar
models and isochrones, and different WD cooling models. It is
instructive, however, to try to single out the influence of the
stellar models on the final error. To this aim,
Fig.~\ref{mibastileo}  shows  a  comparison  between the  progenitor  masses
obtained with two different sets of stellar models  and isochrones (BaSTI and
Padua).  In both cases we used S00 WD models. When compared with the
error bars on $M_{\rm i}$ shown in Fig.~\ref{mimfbastisoo},
Fig.~\ref{mibastileo} leads us 
to conclude that systematic uncertainties  from using
different sets of isochrones and stellar models do not contribute appreciably
to the global  error budget on $M_{\rm i}$, as long as the assumptions
concerning the stellar physics are similar.
We emphasize again that
to achieve internal consistency the  same set of isochrones and stellar models
must be 
used to determine the cluster ages and to obtain progenitor masses from the
progenitor lifetimes.

We next study the changes in the IFMR from using different WD models.
Figure~\ref{mimfs00aldo} displays our  reference  IFMR  (S00  +  BaSTI  with
overshooting) and that resulting from using the LPCODE WD models.  Although WD
masses derived from both sets of WD models are very
similar, the effect  on  the progenitor  mass  is much  larger, especially  for
progenitor masses above $\sim$5~M$_\odot$. The changes are due to differences
in the WD cooling ages obtained with the different WD models. Both S00
and LPCODE models give WD ages consistent within  the observational errors in
\loggs and \teff, however the short lifetime of massive stars
makes the determination of $M_{\rm i}$ very sensitive to small differences in
$t_{\rm cool}$ for higher progenitor masses.

Figure~\ref{fracsigmas} gives an overview of the relative contributions of
observational and model (cooling tracks, including systematic effects due to
variations of input physics and chemical stratification) uncertainties
to the total error budget for the final WD masses and ages. We display also
the relative contribution of cluster age (due to
isochrones, cluster metallicity, reddening, distance) and WD age
(due to cooling tracks) uncertainties to the
total error budget for the progenitor masses and ages.
These fractional contributions are expressed
in terms of $(\sigma_{\rm i}/\sigma_{\rm tot})^2$ for each WD,
denoted by its number (as in Table~\ref{WDsample}), whilst vertical dashed
lines separate the various clusters. The uncertainty of the WD mass
appears to be completely dominated by the observational errors
in \loggs and \teff, while in case of WD ages
the systematic uncertainties of the cooling tracks dominate for some
clusters (NGC3532 and NGC2099). For the progenitor ages and masses (lower panels
of Fig.~\ref{fracsigmas})
cluster ages are the largest contributors to the total error, except mainly for
NGC2099.

In  closing this section, we  comment on the  possibility that systematic 
uncertainties (to be added to the quoted errors) may  be present in the empirical determination  of WDs \teff\ 
and \logg  \citep[see, e.g.,][]{ngs99}. We have not included this source of
uncertainty in our global error assessment simply because it is not quantified
in the literature. We have performed, however, a simple test to understand how 
such hypothetical uncertainties may  influence the determinations of WD masses
and 
cooling ages. We have considered  that \teff\ determinations  have systematic
uncertainties amounting up to $\pm1000$~K, that we represent by means of a 
uniform distribution.  Analogously, in the case of \logg\ we have assumed an
additional  systematic uncertainty of $\pm0.15$~dex.  
Note that  these are  probably very
generous systematic errors, because we assume all WDs are similarly affected.
Our results  show that  uncertainties in \logg\  have a more  noticeable effect
than those on \teff. This is  because WD masses are more robustly predicted by
WD  models  than  cooling  ages,  and consequently  the  error budget  is
dominated    by   observational   uncertainties    (as   already    shown   in
Fig.~\ref{fracsigmas}).  For about  40\%  of our  sample,  uncertainties in  WD
masses  are  about a  factor  of 2  larger  when  systematic uncertainties  as
described  above are  included; these  are the  stars that  show  the smallest
observational  uncertainties in  Table~\ref{WDsample}.   For the  rest of  the
stars in  our sample, uncertainties are increased by not  more than a
factor of 1.5.  Cooling 
ages,   on  the   other   hand,   are  much   less   affected  by   systematic
uncertainties, because the total error  budget here is heavily influenced by
input physics  and systematics  in WD models.  As a consequence,  initial mass
determinations are only very mildly affected.  Indeed, the largest increase in
positive  uncertainties for $M_i$  is only  50\%, with  only 4  objects having
changes 
above 40\%. There are 40  objects for which positive uncertainties increase by
less than  25\%.  Negative uncertainties,  which are always smaller,  are less
affected,  and for  49  objects, increases  are  smaller than  25\%.  We  thus
conclude 
that adding systematic uncertainties to WD observational parameters would only
have a very modest influence on  the inferred IFMR. Based on these results and
also considering that any attempt to quantify systematic uncertainties remains,
at this point, arbitrary, we have  chosen not to include these effects neither
in our reference IFMR nor in the discussion of our results. 


\subsection{Comparison between predicted and derived IFMRs}

Figure~\ref{mimfbastisoo} compares the inferred IFMR with that predicted by
the same BaSTI models (we chose to display the BaSTI results for [Fe/H]=0.06)
used in the determination of $t_{prog}$ and $M_{\rm i}$. We display results
obtained from progenitor models and  isochrones with and without MS
core overshooting.

Cluster ages determined from models without
overshooting are smaller (see
Table~\ref{Ages}),  hence $t_{\rm  prog}$  is decreased,  resulting in  larger
progenitor masses.
 
As a consequence, 
although the associated error bars are large,
models without overshooting predict for some stars  
progenitors masses well above the maximum possible values allowed 
by theory. Examples are the WD in the Pleiades and two of the WDs in
NGC2516 (see Table~\ref{MiMftab}),  
for which the 
estimated progenitor masses are between $\sim 12$ and $\sim 16\,M_{\odot}$. 
Theoretical stellar evolution models predict an upper limit 
to the progenitors of carbon/oxygen white dwarfs in the range 
between $\sim$ 6  and $\sim$ 8 $M_{\odot}$, the exact  values depending on the
chemical 
composition of the models and on the treatment of mixing in the cores; 
lower  upper limits  are obtained  when core  overshooting during  the central
H-burning 
phase is included.  In several other cases in Table~\ref{MiMftab} even negative
progenitor ages were obtained.

The relationship  between $M_{\rm i}$  and the core  mass at the onset  of the
thermal 
pulses ($M_{\rm c1TP}$) is also displayed in Fig.~\ref{mimfbastisoo}.
For models without core overshooting,
the theoretical IFMR lies
largely above  the semi-empirical determination,  especially for $M_{\rm  i} <
5\,M_\odot$, 
while  the theoretical $M_{\rm i}-M_{\rm c1TP}$ relationship 
is in  better agreement with  the data. This  would imply that  the progenitor
stars 
did not experience any thermal pulses at all.

On the other hand, for
core  overshooting  models  the  theoretical  IFMR  follows  much  better  the
semi-empirical
results, and no very high initial masses are typically derived in this case. 
The theoretical $M_{\rm  i}-M_{\rm c1TP}$  relation constitutes a lower envelope to the 
data, consistent with general expectations about AGB evolution.
This would lead us to conclude that core overshooting during the
MS results in a better agreement with the derived IFMR,
because of the internal consistency about the end-product of
stars with inferred $M_{\rm i}$.

Before moving forward in our comparison with theory, we wish to comment
briefly on NGC2099, that represents an interesting case and a warning about
the role played by the cluster [Fe/H] and E(B-V) estimates. As already
mentioned, we have considered two possible values for its metallicity and
reddening. Results for both choices are given in Table~\ref{MiMftab} and
are  also plotted  in Fig.~\ref{fig:m37}.  It becomes apparent from the
figure that data points fall systematically below even the  theoretical
$M_{\rm i}-M_{\rm c1TP}$  (models and IFMR with overshooting) relation for
[Fe/H]= 0.09. The same  result is obtained if models without core
overshooting are used instead.  This inconsistency is strongly alleviated
if    the   alternative    [Fe/H]    and   E(B-V)    used   by    \citet[][see
  Table~\ref{Ages}]{kal05} are 
adopted. While the aim
of this paper is not to use the IFMR to improve cluster properties, this
example demonstrates how informative it can be, when systematic effects
are well controlled.

Focusing again on the core overshooting IFMR (Fig.~\ref{mimfbastisoo},
top panel) we can finally try  to obtain some constraints on TP-AGB evolution
by comparing  the  inferred  IFMR  with  that resulting  from  sequences  with
different TP-AGB modelling.  To do this, we first compare in
Fig.~\ref{mimfbastisoo} the
$M_{\rm i}-M_{\rm  c1TP}$ relations  from  BaSTI and  LPCODE  stellar models  (thick
dotted and  thin dashed lines  respectively). Both relations are  very similar,
regardless of  differences in  the treatment of convective  and additional
mixing episodes. We recall  that  in  BaSTI  instantaneous mixing  in  the
overshooting  region  beyond  the  Schwarzschild boundary  of  the  H-burning
convective core is  assumed, as well as semiconvective  mixing during He-core
burning,  which  is briefly  described  in  \citet{pietr04},  where also  the
appropriate  references are given.   On the  other hand,  no extra  mixing is
taken into account  at the bottom of convective envelopes.
LPCODE  models  include  diffusive  mixing  due  to  overshooting  at  all
convective boundaries. We also point out that evolutionary timescales until
the onset  of the TPs are very  similar in BaSTI models  with overshooting and
the  LPCODE  models.   Since  TP-AGB  lifetimes  are  very   short  and  their
contribution to 
the  the   total  progenitor  age  can   be  neglected,  we   can  safely  use
semi-empirical data 
obtained from BaSTI models for comparisons with the LPCODE theoretical IFMR.

Having assessed  the  similarity  of the  $M_{\rm i}-M_{\rm  c1TP}$
relations, differences in the IFMRs predicted by BaSTI and LPCODE models
can be traced to differences during the TP-AGB phase. While
BaSTI models do not include extra mixing during the thermal pulses, LPCODE
includes, as mentioned above, diffusive overshooting at all convective
boundaries. The theoretical IFMRs are also shown in
Fig.~\ref{mimfbastisoo} with  thick (BaSTI) and thin (LPCODE)  solid lines
and do indeed differ. This is unrelated to the choice of mass loss during
the AGB phase, and it is mainly due to the presence of very strong third
dredge-up episodes in LPCODE sequences, which inhibit the growth of the
H-free core for models with $M_{\rm i} \gtsim 3 {\rm M_\odot}$. These
strong  third dredge-up  episodes  can be  traced  back  mainly to  the
presence of overshooting at the base of the thermal pulse driven
convective zone. This was demonstrated by \cite{Herwig2000} and confirmed
by us running test models where convective overshooting was included only
at separately selected convective boundaries during the AGB evolution. Due
to strong third dredge-up events, the final mass of LPCODE sequences with
$M_{\rm i} \gtsim 3 M_\odot$ is very similar to $M_{\rm c1TP}$, the
core mass at the first thermal pulse (Fig.~\ref{mimfbastisoo}, top panel).
On the other hand in BaSTI sequences the final mass is mainly set by the
length of the TP-AGB phase and, thus, by the adopted mass loss rate. The
apparent failure of the theoretical LPCODE IFMR to match the final masses
in the range $2.75 M_\odot \lessim M_{\rm i}\lessim 3.5 M_\odot$ is an
indication that such strong dredge-up episodes do not take
place in real stars, and therefore the adopted exponentially decaying
overshooting at the base of the thermal pulse driven convective zone in
the LPCODE models  is too  strong.  This  allows us  to conclude  that the
overshooting parameter $f$ \citep[see][for a definition of
$f$]{Herwig1997} must be significantly smaller than $f=0.016$ employed at
the convective core of upper MS stars. This is in line with
recent hydrodynamical simulations of AGB thermal pulses which point to a
value of $f\lessim 0.01$ at the bottom of the pulse driven convective zone
\citep{Herwig2007}. On the other hand, given the large error bars of the
data points, the existence of strong third dredge-up episodes for stars
with $M_{\rm i}\gtsim 5 M_\odot$ cannot be excluded by the same
arguments. The IFMR may therefore teach us that overshooting varies a lot
between different convective layers and may also depend on stellar mass.

\section{Summary and conclusions}

Since its earliest determination, the semi-empirical IFMR has
indicated a monotonic rise of the final WD mass with the initial mass
of the progenitor. It is also strong evidence that single stars of
intermediate mass do not develop degenerate cores close to the
Chandrasekhar limit, most likely due to the effect of strong mass loss
on the AGB. These qualitative results have remained unchanged
in spite of the significant improvement both in the amount and
accuracy of observational data and in the theoretical
models. Nevertheless there are still many open questions, such as a
dependence on metallicity, the slope of the IFMR at the upper
(initial) mass end, changes in slope, and spread around a mean
relation. To answer these questions, not only increasingly
accurate observations are needed, but also the systematic
uncertainties coming from the theoretical input -- which is
necessary to establish the semi-empirical IFMR --  have to be
controlled. These are related to
evolutionary tracks and isochrones that determine the total age of the
observed WDs and cooling tracks used for deriving the WD
age. Our paper was intended to investigate the influence of such
systematic uncertainties and to derive robust features of the IFMR. We
also investigated what uncertainties in the theory of stellar
evolution would lead to inconsistent results, and therefore could be
restricted or excluded.

From the available WD data we selected a total of 52 objects in 10 open
clusters, plus Sirius B. There are, in fact, even more objects
available, but we disregarded those for which our cluster distance
determination (Sect.~2) would not be possible because of the lack of
calibrating field stars at that metallicity (NGC6791 at ${\rm [Fe/H]}
= 0.45$). We also did not take into account the objects from
\citet{cata07}, because of the more indirect  way to obtain the age
of the mostly unevolved binary companion. Inclusion of
such objects would weaken our aim of highest possible
self-consistency.

For the determination of mass and cooling age of the WDs we
used different sets of cooling track, from \citet{sala00}
and the LPCODE \citep{alt03} to evaluate the influence of
different codes, input physics and chemical stratification
(Sect.~3.1). We estimated the error ranges quite
conservatively and to evaluate the global error budget
we used Monte Carlo simulations (Sect.~3.3), for
which we assumed either a normal or uniform distribution of the
different error sources. We included the effect of observational errors, too.

Results concerning cooling ages and final masses are quite
robust: the accuracy of the WD masses depend completely on the
observational uncertainties, while their ages is influenced in most
cases (in terms of number of clusters) by the
uncertainties of the cooling tracks, mostly systematic differences between
different sets of cooling tracks and changes in the neutrino emission rates.

\begin{figure}
\epsscale{1.00}
\plotone{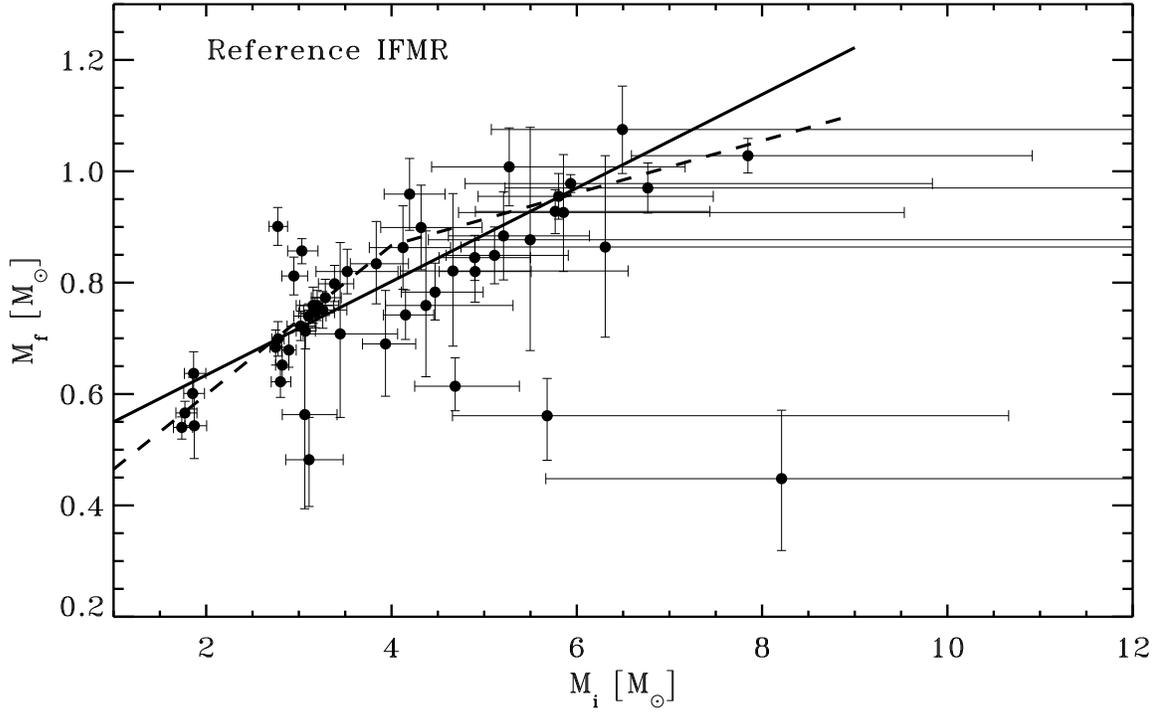}
\caption{Reference IFMR obtained employing BaSTI isochrones and progenitor
 models with core overshooting together with S00 WD tracks. Lines show fits to
 the data: the solid line  is a linear  fit, the dashed  line a piecewise linear fit.
 The pivot point at 4~M$_\odot$ has been chosen
 following theoretical predictions from BaSTI models, that show changes in the
 slope of the IFMR around those mass values.
\label{fig:ref_ifmr} }
\end{figure}

This result implies that any spread in $M_{\rm f}$, as seen in
Fig.~\ref{fig:ref_ifmr}, which is outside the error bar is real, in the
sense that $M_{\rm f}$ does vary at the same (within the errors in $M_{\rm
i}$) initial mass. Whether this spread is consistent with theoretical
expectations, related to a steep increase of $M_{\rm f}$ with $M_{\rm i}$
in that mass range, or whether it is due to star-to-star variations in the
total mass lost, cannot be decided yet, as the error bars in $M_{\rm i}$
are too large even in the case of the Praesepe WDs, which we discussed in
detail in Sect.~2.4.

Figure~\ref{mimfs00aldo} demonstrates that uncertainties due
to different WD codes
with similar, but not identical physics and model details are essentially
irrelevant  for the  estimate  of  $M_{\rm f}$.  However,  differences in  the
cooling ages 
have a significant influence on $M_{\rm i}$, since they affect directly
progenitor lifetimes, and the more massive progenitors evolve on
very short timescales. Therefore the two different sets of cooling
tracks lead  to quite  different values for  $M_{\rm i}$, particularly for
initial masses above 5~M$_\odot$.  
In  this  regime, $M_{\rm  i}$  differences are  between  $\sim$  0.5 and  2.0
$M_{\odot}$.  

The determination of $M_{\rm i}$ is more complex, as it involves several
steps. The age determination of the cluster or binary system requires
metallicity, reddening and distance. We have used literature values for the
former two, and our own homogenous determination for the
latter. Uncertainties from both steps have been taken into
account. Using appropriate isochrones the age is then
obtained, which, together with the WD cooling age, results in the
pre-WD lifetime. As this is, to an accuracy of 1\% or slightly more,
identical to  the lifetime from  the MS to  the end of the  central He-burning
phase, 
all uncertainties concerning the AGB evolution are
irrelevant. We employed different sets of stellar models
\citep{pietr04,gir00}, which differ slightly in the detailed treatment
of the input physics (but both include core overshooting during the MS)
as well as additional models from the
former source (BaSTI), which do not consider convective
overshooting. The first effect -- different tracks -- has a very minor
effect on $M_{\rm i}$, as we showed in Fig.~\ref{mibastileo}. Note that this
includes both the cluster age determination and the relation between
progenitor age and mass. The neglect of overshooting, however, leads
to internal inconsistencies in the semi-empirical IFMR: at the high
end, initial masses are predicted, for which the models themselves
predict that no electron degenerate CO-WD would result from the evolution.
By discriminating
between the different error sources (but omitting the overshooting
effect), Fig.~\ref{fracsigmas} demonstrates that the cluster age
uncertainties  (due  to uncertainties  in  the  cluster  [Fe/H] and  reddening
estimates) 
dominate the error on $M_{\rm i}$, although the above-mentioned WD age
can be significant for the more massive progenitors, too.

The case of NGC2099 illustrates the importance of
accurate metallicities and reddenings:
Figure~\ref{fig:m37} shows how drastically the semi-empirical IFMR
changes, two different estimates of [Fe/H] and $E(B-V)$ are assumed.
If one accepts this uncertainty as being real, NGC2099
cannot be used at all to learn about the IFMR.

\begin{figure}
\epsscale{0.9}
\plotone{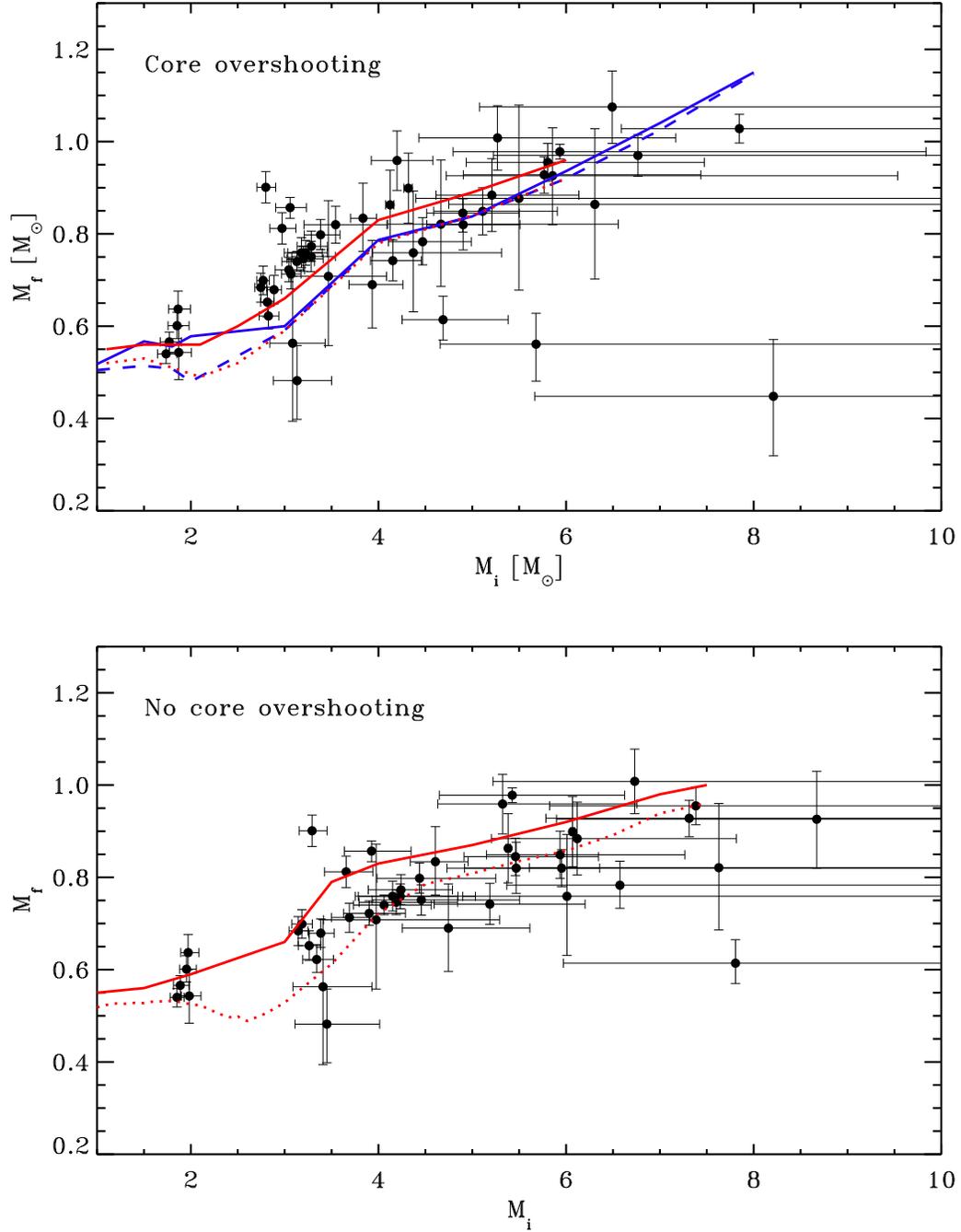}
\caption{Inferred  and theoretical IFMRs.   Top panel:  results for  models and
  isochrones  with core  overshooting. Solid  lines correspond  to theoretical
  IFMRs derived from BaSTI models  (thick lines -- for [Fe/H]=0.06) and LPCODE
  models
(thin  line --  for   [Fe/H]=  0.06). The  dotted thick  line shows  the
  $M_{\rm i}-M_{\rm c1TP}$ relation for BaSTI
models  and the dashed  thin line  shows the counterpart from LPCODE models.
  Bottom panel:  results corresponding to  models and isochrones  without core
  overshooting. In  this case  only theoretical results  for BaSTI  models are
  shown.  \label{mimfbastisoo}}
 \end{figure}

\begin{figure}
\epsscale{1.00}
\plotone{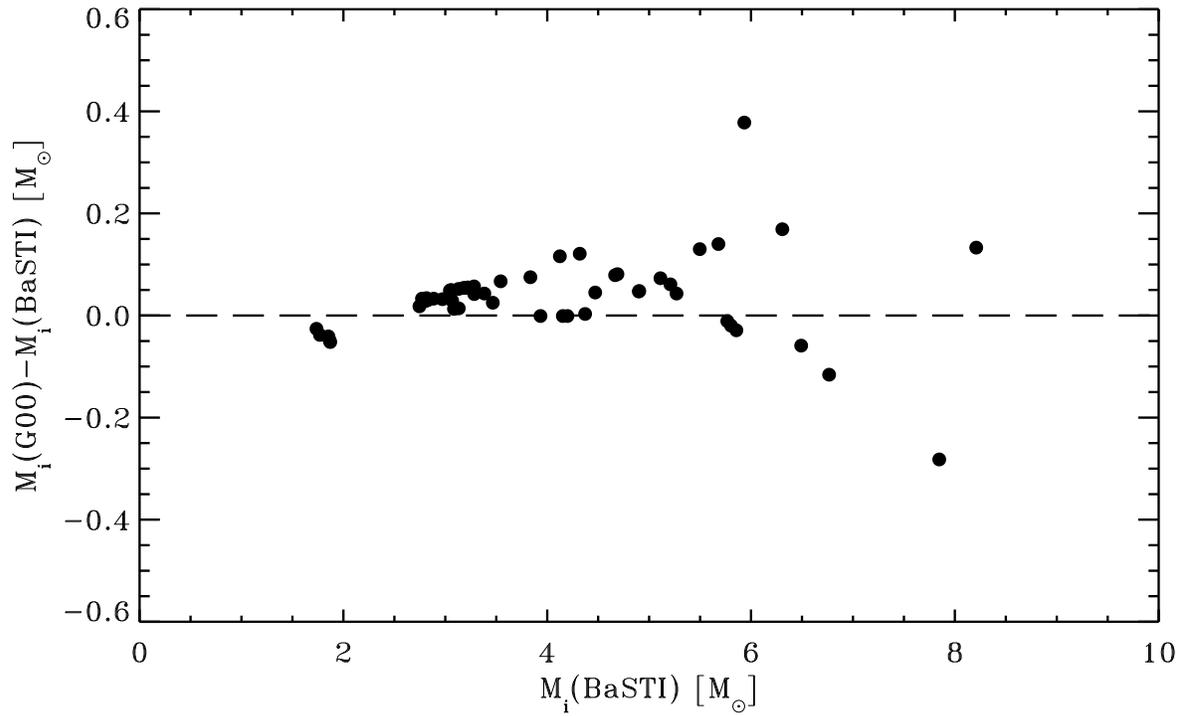}
\caption{Difference  between  initial  masses obtained  employing
  the  Padua and 
  BaSTI  isochrones  and models  versus  initial  masses  obtained with  BaSTI
  isochrones and models. Both sets
  include overshooting treated with different formulations, but giving
  similar  mass extensions  of the  core overshooting  regions at  fixed total
  stellar    mass.     S00   WD    tracks    have    been    used   in    both
  cases. \label{mibastileo}} 
\end{figure}

\begin{figure}
\epsscale{1.00}
\plotone{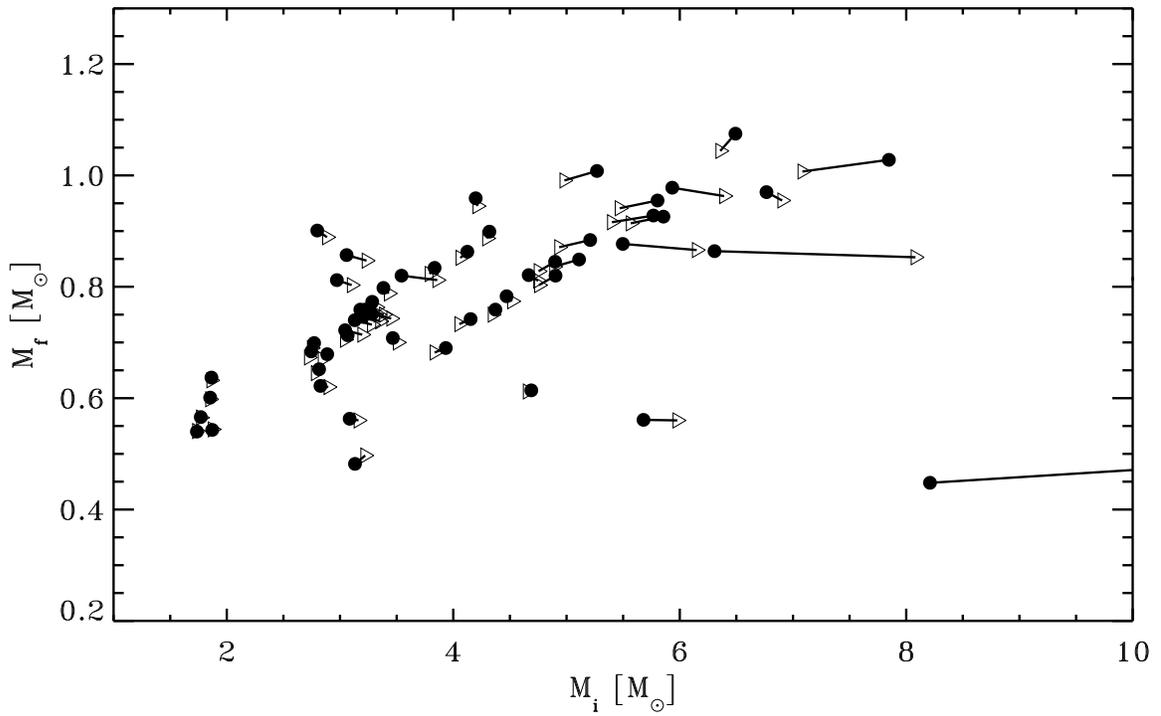}
\caption{Comparison between IFMRs obtained using the S00 (dots) and the
 LPCODE (triangles) WD tracks. BaSTI isochrones and progenitor models with
 core overshooting have been used in both cases. Each pair connected by a line
 represents the same star.
\label{mimfs00aldo}}
\end{figure}

\begin{figure}
\epsscale{1.00}
\plotone{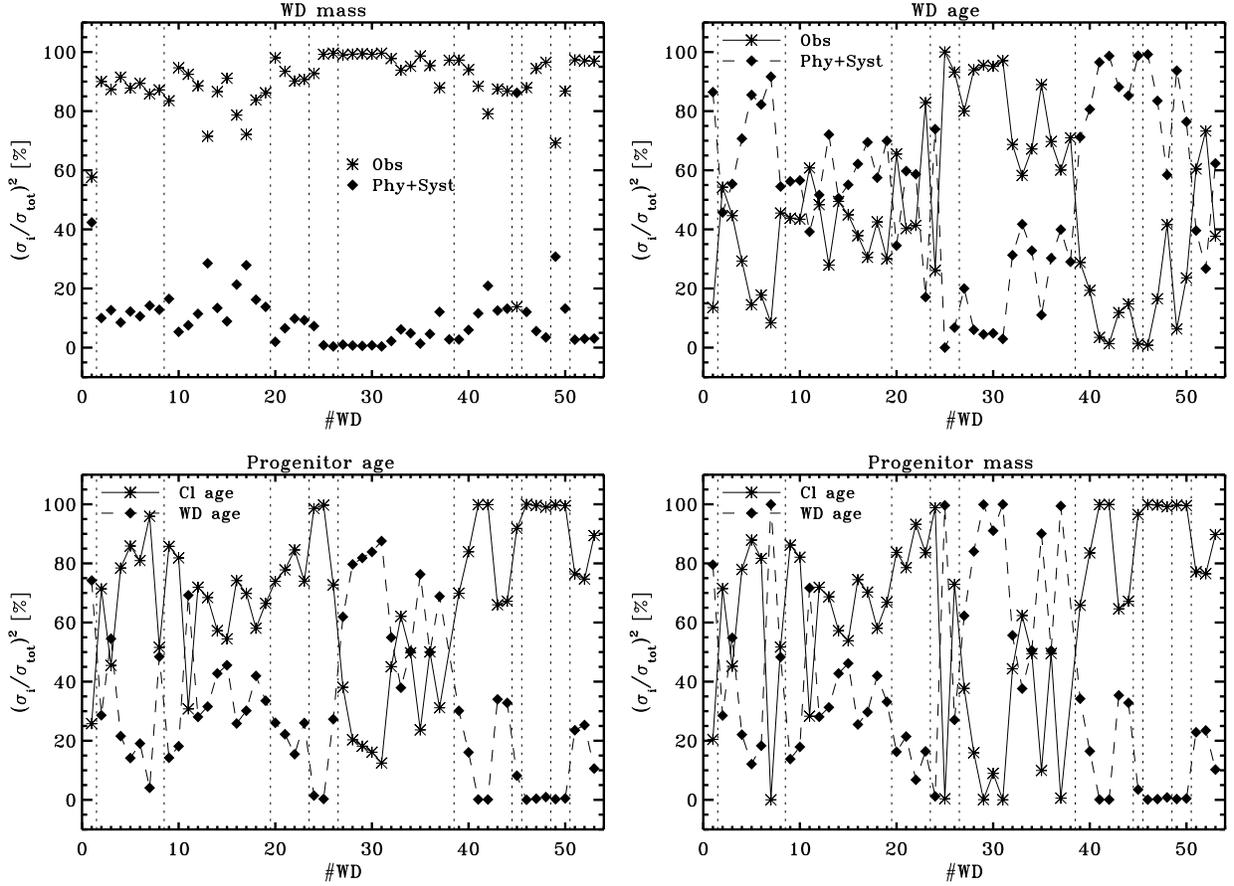}
\caption{Contribution of uncertainty sources to the total error budget
  for WD mass (upper left), WD age (upper right), progenitor mass
  (lower  left),  and  progenitor  age  (lower right  panel).  The  fractional
  contribution to total sigma
  $\sigma$ due to various error sources is given for each WD. Vertical
  lines separate clusters. For the WDs we show the importance of
  observational vs.\ input physics and systematics (different codes, variation
  of physics inputs,
  chemical stratification, envelope thickness) for the
  progenitors that of cluster and WD age. \label{fracsigmas}}
\end{figure}

\begin{figure}
\epsscale{1.00}
\plotone{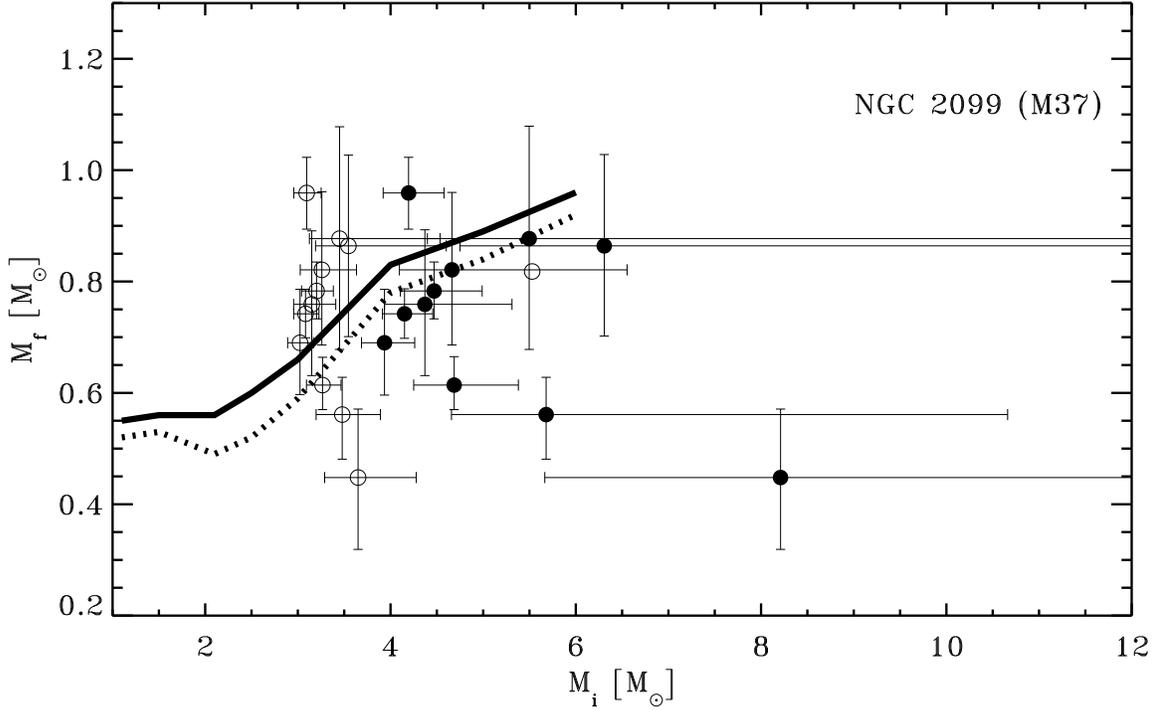}
\caption{IFMR for NGC 2099, for the two [Fe/H] and E(B-V) pairs
  given in Table~\ref{Ages}. Filled  circles  refer  to  [Fe/H]= 0.09  and  empty  circles  to  [Fe/H]=
  $-$0.20. The solid line shows the theoretical IFMR from BaSTI models with core
  overshooting, the dotted line displays the $M_{\rm i}-M_{\rm c1TP}$  relation for
  the same models. Data  points lie systematically below the
  theoretical $M_{\rm i}-M_{\rm c1TP}$ values when [Fe/H]=  0.09 is adopted. \label{fig:m37}}
\end{figure}

After we had determined the full extent of the uncertainty associated
to our semi-empirical IFMR, we compared it with the predictions from the
theoretical calculations, both from BaSTI and LPCODE. Both differ in
their treatment of convective overshooting during the AGB phase. One
should recall that the BaSTI models treat that phase with synthetic
models, while LPCODE follows the evolution in full
detail. Figure~\ref{mimfbastisoo} shows them in comparison with the
semi-empirical IFMR and the relation between $M_{\rm i}$ and the mass of the
CO-core at the first thermal pulse, $M_{\rm c1TP}$. As long as stars
do not avoid the AGB phase, $M_{\rm f} \geq M_{\rm c1TP}$ should be
fulfilled. Therefore, this relation is a strong lower limit to the
semi-empirical IFMR. The agreement between the $M_{\rm i}-M_{\rm c1TP}$
relations from both codes is excellent;
there are no
significant systematic uncertainties in its prediction. This again
emphasizes the need for convective overshooting during the
MS phase:  without it a significantly larger numbers of
objects falls below this limiting line.

On the other hand, the theoretical IFMR from the LPCODE tracks is
not consistent with the data: In these models overshooting during
the AGB, mainly from the lower boundary of the pulse-driven convective
zone, limits the growth of the CO-core such that the final core size
is almost identical to that at the first pulse. However, most objects
have final masses clearly above this line, such that we concluded that
the extent of overshooting during the AGB phase must be significantly
smaller than on the MS. This is in agreement with
hydrodynamical simulations.

The   theoretical  IFMRs  displayed   in  Figure~\ref{mimfbastisoo},   and  in
particular those 
resulting from the BaSTI models display the largest gradient in the same mass
range where a significant spread in $M_{\rm f}$ is observed. We demonstrated
that indeed that spread could arise solely from this local steep gradient, such
that no other explanation would be needed. If it is not, one would have to
invoke a spread in the total mass lost in a given, small mass range. This is not
exceptional, as the morphology of horizontal branches
in Galactic globular cluster shows, but the reasons for
such a star-to-star variation are unclear. In other mass ranges the total error
budget is too large to discuss the reality and reason for apparent spreads in
$M_{\rm f}$.

Figure~\ref{fig:ref_ifmr} presents our reference semi-empirical IFMR, obtained
using BaSTI isochrones and tracks with overshooting and
S00 cooling models. The error bars include all errors we have
investigated. Compared to \citet{fer05} our error bars are necessarily larger
(we consider more sources of error), but
on average only by  less than a factor of two, and  mostly for $M_{\rm i}$. We
also present a 
simple linear fit through all the data
as well as a piecewise fit, which reflects the
properties of the theoretical IFMR based on the BaSTI models.

In summary, our most important results are
\begin{enumerate}
\item None of the WDs employed in current IFMR determinations is close to the
 Chandrasekhar mass, not even the progeny of the more massive
 intermediate mass stars;
\item stellar models without convective overshooting during core hydrogen
  burning lead to internal inconsistencies in the semi-empirical IFMR;
\item overshooting from convective boundaries during the AGB phase must be
  significantly reduced (compared to the case of convective cores along the MS)
  to reproduce the observed $M_{\rm f}$ values;
\item the uncertainty in $M_{\rm f}$ is dominated by observational errors;
\item  the  uncertainty in  $M_{\rm  i}$  has  several reasons:  both  cluster
  parameters and 
  isochrone details influence the cluster age and thus the progenitor mass; the
  uncertainty on the WD cooling age can sometimes also be the dominant factor;
\item the observed dispersion in $M_{\rm f}$ at approximately constant $M_{\rm
  i}$ (3.0--3.5 $M_{\odot}$), 
  in particular for the Praesepe objects, appears to be real. It
  may follow from the steep increase of $M_{\rm f}$ with $M_{\rm i}$ predicted
  by theoretical IFMRs 
  in that $M_{\rm i}$ range, rather than being caused by a spread in
  mass loss among cluster AGB stars;
\item from the general agreement  between the theoretical  IFMRs and the
  semiempirical data, 
  we find no evidence that the mass loss prescriptions used in the
  stellar evolution calculations grossly disagree with the total mass
  actually lost by low- and intermediate- mass stars;
\item  the case  of NGC2099  illustrates  the necessity  for accurate  cluster
  parameters; 
  reliable composition data and reddening are required;
\item as long as up-to-date input physics is used for the stellar models,
  systematic uncertainties do not change the overall appearance of the
  semi-empirical IFMR as determined recently by several groups.
\end{enumerate}

An extension of this analysis to a larger range of [Fe/H] is needed,
to investigate a possible metallicity dependence of the IFMR, given
the small [Fe/H] range spanned by the systems analyzed in this study
(see Table~\ref{Ages}).  In addition to the relevant observational
data, this requires also a careful study of how to extend to a wider
[Fe/H] range the empirical MS fitting technique adopted
here to determine cluster distances, if we wish to maintain the same
degree of homogeneity in the derivation of the cluster ages.  
An extension to
lower metallicities seems 
to be preferable, as both calibrating objects and suitable clusters
should exist, contrary to the case of an extension to higher, that is
super-solar clusters.

As most of the clusters used for deriving the IFMR have nearly solar
metallicities, the question arises, what the true solar chemical
composition is? \citet{ags:2005} have presented new determinations of
the solar heavy element abundance, drastically lower than the standard
composition assumed here and in the stellar evolution tracks
employed. Both BaSTI and LPCODE progenitor models consider 
a standard solar mixture with $Z/X \sim$0.0245, whereas 
\citet{ags:2005} redetermination gives $Z/X \sim$0.0165.
Assuming that the differential abundance analyses for
our open cluster sample provides the true abundances relative to the sun, also
the clusters should be less metal-rich than thought. It would be
interesting to see how this will affect the semi-empirical IFMR and
its agreement with theoretical predictions. An extension of this
investigation by using  in the computation of progenitor  models the new solar
abundances is currently under way.

\acknowledgments

We  wish to  thank  our referee,  K.   Williams, for  several suggestions that  helped
improving this  work. We are grateful  to J.~Kalirai for pointing  out his new
results to us and for continuing interest in our project. 
AMS was supported by the IAS through a John Bahcall Fellowship and the NSF
through the grant PHY-0503584. M3B is supported by CONICET through a doctoral
fellowship.

\bibliographystyle{apj}
\bibliography{bib_MiMf}

\begin{thebibliography}{79}
\expandafter\ifx\csname natexlab\endcsname\relax\def\natexlab#1{#1}\fi

\bibitem[{{Althaus} {et~al.}(2003){Althaus}, {Serenelli}, {C{\'o}rsico}, \&
  {Montgomery}}]{alt03}
{Althaus}, L.~G., {Serenelli}, A.~M., {C{\'o}rsico}, A.~H., \& {Montgomery},
  M.~H. 2003, \aap, 404, 593

\bibitem[{{Asplund} {et~al.}(2005){Asplund}, {Grevesse}, \&
  {Sauval}}]{ags:2005}
{Asplund}, M., {Grevesse}, N., \& {Sauval}, A.~J. 2005, in Astronomical Society
  of the Pacific Conference Series, Vol. 336, Cosmic Abundances as Records of
  Stellar Evolution and Nucleosynthesis, ed. T.~G. {Barnes}, III \& F.~N.
  {Bash}, 25--+

\bibitem[{{Barstow} {et~al.}(2005){Barstow}, {Bond}, {Holberg}, {Burleigh},
  {Hubeny}, \& {Koester}}]{bars05}
{Barstow}, M.~A., {Bond}, H.~E., {Holberg}, J.~B., {Burleigh}, M.~R., {Hubeny},
  I., \& {Koester}, D. 2005, \mnras, 362, 1134

\bibitem[{{Bischoff-Kim} {et~al.}(2008){Bischoff-Kim}, {Montgomery}, \&
  {Winget}}]{bish08}
{Bischoff-Kim}, A., {Montgomery}, M.~H., \& {Winget}, D.~E. 2008, \apj, 675,
  1512

\bibitem[{{Bragaglia} {et~al.}(2001){Bragaglia}, {Carretta}, {Gratton}, {Tosi},
  {Bonanno}, {Bruno}, {Cal{\`i}}, {Claudi}, {Cosentino}, {Desidera},
  {Farisato}, {Rebeschini}, \& {Scuderi}}]{brag01}
{Bragaglia}, A., {Carretta}, E., {Gratton}, R.~G., {Tosi}, M., {Bonanno}, G.,
  {Bruno}, P., {Cal{\`i}}, A., {Claudi}, R., {Cosentino}, R., {Desidera}, S.,
  {Farisato}, G., {Rebeschini}, M., \& {Scuderi}, S. 2001, \aj, 121, 327

\bibitem[{{Cassisi} {et~al.}(2007){Cassisi}, {Potekhin}, {Pietrinferni},
  {Catelan}, \& {Salaris}}]{cass07}
{Cassisi}, S., {Potekhin}, A.~Y., {Pietrinferni}, A., {Catelan}, M., \&
  {Salaris}, M. 2007, \apj, 661, 1094

\bibitem[{{Castanheira} \& {Kepler}(2008)}]{ck08}
{Castanheira}, B.~G. \& {Kepler}, S.~O. 2008, \mnras, 385, 430

\bibitem[{{Catal{\'a}n} {et~al.}(2008{\natexlab{a}}){Catal{\'a}n}, {Isern},
  {Garc{\'{\i}}a-Berro}, \& {Ribas}}]{cata08}
{Catal{\'a}n}, S., {Isern}, J., {Garc{\'{\i}}a-Berro}, E., \& {Ribas}, I.
  2008{\natexlab{a}}, \mnras, 387, 1693

\bibitem[{{Catal{\'a}n} {et~al.}(2008{\natexlab{b}}){Catal{\'a}n}, {Isern},
  {Garc{\'{\i}}a-Berro}, {Ribas}, {Allende Prieto}, \& {Bonanos}}]{cata07}
{Catal{\'a}n}, S., {Isern}, J., {Garc{\'{\i}}a-Berro}, E., {Ribas}, I.,
  {Allende Prieto}, C., \& {Bonanos}, A.~Z. 2008{\natexlab{b}}, \aap, 477, 213

\bibitem[{{Chen} {et~al.}(2003){Chen}, {Hou}, \& {Wang}}]{chen03}
{Chen}, L., {Hou}, J.~L., \& {Wang}, J.~J. 2003, \aj, 125, 1397

\bibitem[{{Claver} {et~al.}(2001){Claver}, {Liebert}, {Bergeron}, \&
  {Koester}}]{clav01}
{Claver}, C.~F., {Liebert}, J., {Bergeron}, P., \& {Koester}, D. 2001, \apj,
  563, 987

\bibitem[{{Deliyannis} {et~al.}(2002){Deliyannis}, {Jacobson}, {Cummings},
  {Steinhauer}, \& {Sarajedini}}]{del02}
{Deliyannis}, C.~P., {Jacobson}, H., {Cummings}, J., {Steinhauer}, A., \&
  {Sarajedini}, A. 2002, in Bulletin of the American Astronomical Society,
  Vol.~34, Bulletin of the American Astronomical Society, 1308--+

\bibitem[{{Dobbie} {et~al.}(2006{\natexlab{a}}){Dobbie}, {Napiwotzki},
  {Burleigh}, {Barstow}, {Boyce}, {Casewell}, {Jameson}, {Hubeny}, \&
  {Fontaine}}]{dobbie06}
{Dobbie}, P.~D., {Napiwotzki}, R., {Burleigh}, M.~R., {Barstow}, M.~A.,
  {Boyce}, D.~D., {Casewell}, S.~L., {Jameson}, R.~F., {Hubeny}, I., \&
  {Fontaine}, G. 2006{\natexlab{a}}, \mnras, 369, 383

\bibitem[{{Dobbie} {et~al.}(2006{\natexlab{b}}){Dobbie}, {Napiwotzki},
  {Lodieu}, {Burleigh}, {Barstow}, \& {Jameson}}]{doetal06}
{Dobbie}, P.~D., {Napiwotzki}, R., {Lodieu}, N., {Burleigh}, M.~R., {Barstow},
  M.~A., \& {Jameson}, R.~F. 2006{\natexlab{b}}, \mnras, 373, L45

\bibitem[{{Dobbie} {et~al.}(2004){Dobbie}, {Pinfield}, {Napiwotzki}, {Hambly},
  {Burleigh}, {Barstow}, {Jameson}, \& {Hubeny}}]{dob04}
{Dobbie}, P.~D., {Pinfield}, D.~J., {Napiwotzki}, R., {Hambly}, N.~C.,
  {Burleigh}, M.~R., {Barstow}, M.~A., {Jameson}, R.~F., \& {Hubeny}, I. 2004,
  \mnras, 355, L39

\bibitem[{{Dominguez} {et~al.}(1996){Dominguez}, {Straniero}, {Tornambe}, \&
  {Isern}}]{domi96}
{Dominguez}, I., {Straniero}, O., {Tornambe}, A., \& {Isern}, J. 1996, \apj,
  472, 783

\bibitem[{{Fernandez} \& {Salgado}(1980)}]{fern80}
{Fernandez}, J.~A. \& {Salgado}, C.~W. 1980, \aaps, 39, 11

\bibitem[{{Ferrario} {et~al.}(2005){Ferrario}, {Wickramasinghe}, {Liebert}, \&
  {Williams}}]{fer05}
{Ferrario}, L., {Wickramasinghe}, D., {Liebert}, J., \& {Williams}, K.~A. 2005,
  \mnras, 361, 1131

\bibitem[{{Fontaine} {et~al.}(2001){Fontaine}, {Brassard}, \&
  {Bergeron}}]{font01}
{Fontaine}, G., {Brassard}, P., \& {Bergeron}, P. 2001, \pasp, 113, 409

\bibitem[{{Gim} {et~al.}(1998){Gim}, {Vandenberg}, {Stetson}, {Hesser}, \&
  {Zurek}}]{gim98}
{Gim}, M., {Vandenberg}, D.~A., {Stetson}, P.~B., {Hesser}, J.~E., \& {Zurek},
  D.~R. 1998, \pasp, 110, 1318

\bibitem[{{Girardi} {et~al.}(2002){Girardi}, {Bertelli}, {Bressan}, {Chiosi},
  {Groenewegen}, {Marigo}, {Salasnich}, \& {Weiss}}]{gir02}
{Girardi}, L., {Bertelli}, G., {Bressan}, A., {Chiosi}, C., {Groenewegen},
  M.~A.~T., {Marigo}, P., {Salasnich}, B., \& {Weiss}, A. 2002, \aap, 391, 195

\bibitem[{{Girardi} {et~al.}(2000){Girardi}, {Bressan}, {Bertelli}, \&
  {Chiosi}}]{gir00}
{Girardi}, L., {Bressan}, A., {Bertelli}, G., \& {Chiosi}, C. 2000, \aaps, 141,
  371

\bibitem[{{Gratton}(2000)}]{grat00}
{Gratton}, R. 2000, in Astronomical Society of the Pacific Conference Series,
  Vol. 198, Stellar Clusters and Associations: Convection, Rotation, and
  Dynamos, ed. R.~{Pallavicini}, G.~{Micela}, \& S.~{Sciortino}, 225--+

\bibitem[{{Gratton} {et~al.}(2006){Gratton}, {Bragaglia}, {Carretta}, \&
  {Tosi}}]{gra06}
{Gratton}, R., {Bragaglia}, A., {Carretta}, E., \& {Tosi}, M. 2006, \apj, 642,
  462

\bibitem[{{Grevesse} \& {Noels}(1993)}]{gn93}
{Grevesse}, N. \& {Noels}, A. 1993, in Origin and Evolution of the Elements,
  ed. S.~{Kubono} \& T.~{Kajino}, 14--+

\bibitem[{{Haft} {et~al.}(1994){Haft}, {Raffelt}, \& {Weiss}}]{haft94}
{Haft}, M., {Raffelt}, G., \& {Weiss}, A. 1994, \apj, 425, 222

\bibitem[{{Herwig}(2000)}]{Herwig2000}
{Herwig}, F. 2000, \aap, 360, 952

\bibitem[{{Herwig} {et~al.}(1997){Herwig}, {Bloecker}, {Schoenberner}, \& {El
  Eid}}]{Herwig1997}
{Herwig}, F., {Bloecker}, T., {Schoenberner}, D., \& {El Eid}, M. 1997, \aap,
  324, L81

\bibitem[{{Herwig} {et~al.}(2007){Herwig}, {Freytag}, {Fuchs}, {Hansen},
  {Hueckstaedt}, {Porter}, {Timmes}, \& {Woodward}}]{Herwig2007}
{Herwig}, F., {Freytag}, B., {Fuchs}, T., {Hansen}, J.~P., {Hueckstaedt},
  R.~M., {Porter}, D.~H., {Timmes}, F.~X., \& {Woodward}, P.~R. 2007, in
  Astronomical Society of the Pacific Conference Series, Vol. 378, Why Galaxies
  Care About AGB Stars: Their Importance as Actors and Probes, ed.
  F.~{Kerschbaum}, C.~{Charbonnel}, \& R.~F. {Wing}, 43--+

\bibitem[{{Iben} \& {Renzini}(1983)}]{iberen83}
{Iben}, Jr., I. \& {Renzini}, A. 1983, \araa, 21, 271

\bibitem[{{Isern} {et~al.}(2008){Isern}, {Garc{\'{\i}}a-Berro}, {Torres}, \&
  {Catal{\'a}n}}]{ise08}
{Isern}, J., {Garc{\'{\i}}a-Berro}, E., {Torres}, S., \& {Catal{\'a}n}, S.
  2008, \apjl, 682, L109

\bibitem[{{Itoh} {et~al.}(1993){Itoh}, {Hayashi}, \& {Kohyama}}]{itoh93}
{Itoh}, N., {Hayashi}, H., \& {Kohyama}, Y. 1993, \apj, 418, 405

\bibitem[{{Itoh} {et~al.}(1996){Itoh}, {Hayashi}, {Nishikawa}, \&
  {Kohyama}}]{itoh96}
{Itoh}, N., {Hayashi}, H., {Nishikawa}, A., \& {Kohyama}, Y. 1996, \apjs, 102,
  411

\bibitem[{{Johnson}(1952)}]{j52}
{Johnson}, H.~L. 1952, \apj, 116, 640

\bibitem[{{Johnson} \& {Knuckles}(1955)}]{j55}
{Johnson}, H.~L. \& {Knuckles}, C.~F. 1955, \apj, 122, 209

\bibitem[{{Johnson} \& {Mitchell}(1958)}]{j58}
{Johnson}, H.~L. \& {Mitchell}, R.~I. 1958, \apj, 128, 31

\bibitem[{{Jones} \& {Prosser}(1996)}]{jonpros96}
{Jones}, B.~F. \& {Prosser}, C.~F. 1996, \aj, 111, 1193

\bibitem[{{Kalirai} {et~al.}(2008){Kalirai}, {Hansen}, {Kelson}, {Reitzel},
  {Rich}, \& {Richer}}]{kalir08}
{Kalirai}, J.~S., {Hansen}, B.~M.~S., {Kelson}, D.~D., {Reitzel}, D.~B.,
  {Rich}, R.~M., \& {Richer}, H.~B. 2008, \apj, 676, 594

\bibitem[{{Kalirai} {et~al.}(2001{\natexlab{a}}){Kalirai}, {Richer}, {Fahlman},
  {Cuillandre}, {Ventura}, {D'Antona}, {Bertin}, {Marconi}, \&
  {Durrell}}]{kalir01}
{Kalirai}, J.~S., {Richer}, H.~B., {Fahlman}, G.~G., {Cuillandre}, J.-C.,
  {Ventura}, P., {D'Antona}, F., {Bertin}, E., {Marconi}, G., \& {Durrell},
  P.~R. 2001{\natexlab{a}}, \aj, 122, 266

\bibitem[{{Kalirai} {et~al.}(2005){Kalirai}, {Richer}, {Reitzel}, {Hansen},
  {Rich}, {Fahlman}, {Gibson}, \& {von Hippel}}]{kal05}
{Kalirai}, J.~S., {Richer}, H.~B., {Reitzel}, D., {Hansen}, B.~M.~S., {Rich},
  R.~M., {Fahlman}, G.~G., {Gibson}, B.~K., \& {von Hippel}, T. 2005, \apjl,
  618, L123

\bibitem[{{Kalirai} {et~al.}(2001{\natexlab{b}}){Kalirai}, {Ventura}, {Richer},
  {Fahlman}, {Durrell}, {D'Antona}, \& {Marconi}}]{kal01a}
{Kalirai}, J.~S., {Ventura}, P., {Richer}, H.~B., {Fahlman}, G.~G., {Durrell},
  P.~R., {D'Antona}, F., \& {Marconi}, G. 2001{\natexlab{b}}, \aj, 122, 3239

\bibitem[{{Koester} \& {Reimers}(1993)}]{kr93}
{Koester}, D. \& {Reimers}, D. 1993, \aap, 275, 479

\bibitem[{{Koester} \& {Reimers}(1996)}]{kr96}
---. 1996, \aap, 313, 810

\bibitem[{{Koester} \& {Weidemann}(1980)}]{koeweid80}
{Koester}, D. \& {Weidemann}, V. 1980, \aap, 81, 145

\bibitem[{{Liebert} {et~al.}(2005){Liebert}, {Young}, {Arnett}, {Holberg}, \&
  {Williams}}]{lieb05}
{Liebert}, J., {Young}, P.~A., {Arnett}, D., {Holberg}, J.~B., \& {Williams},
  K.~A. 2005, \apjl, 630, L69

\bibitem[{{Loktin} {et~al.}(2001){Loktin}, {Gerasimenko}, \&
  {Malishev}}]{lot01}
{Loktin}, A.~V., {Gerasimenko}, T.~P., \& {Malishev}, L.~K. 2001, Astron.
  Astrophys. Trans., 20, 607

\bibitem[{{Marigo} \& {Girardi}(2007)}]{mg07}
{Marigo}, P. \& {Girardi}, L. 2007, \aap, 469, 239

\bibitem[{{Meynet} {et~al.}(1993){Meynet}, {Mermilliod}, \&
  {Maeder}}]{meynet93}
{Meynet}, G., {Mermilliod}, J.-C., \& {Maeder}, A. 1993, \aaps, 98, 477

\bibitem[{{Mochejska} \& {Kaluzny}(1999)}]{mk99}
{Mochejska}, B.~J. \& {Kaluzny}, J. 1999, Acta Astronomica, 49, 351

\bibitem[{{Morel}(1997)}]{morel97}
{Morel}, P. 1997, \aaps, 124, 597

\bibitem[{{Napiwotzki} {et~al.}(1999){Napiwotzki}, {Green}, \&
  {Saffer}}]{ngs99}
{Napiwotzki}, R., {Green}, P.~J., \& {Saffer}, R.~A. 1999, \apj, 517, 399

\bibitem[{{Nilakshi} \& {Sagar}(2002)}]{sag02}
{Nilakshi} \& {Sagar}, R. 2002, \aap, 381, 65

\bibitem[{{Percival} \& {Salaris}(2003)}]{per03b}
{Percival}, S.~M. \& {Salaris}, M. 2003, \mnras, 343, 539

\bibitem[{{Percival} {et~al.}(2005){Percival}, {Salaris}, \&
  {Groenewegen}}]{per05}
{Percival}, S.~M., {Salaris}, M., \& {Groenewegen}, M.~A.~T. 2005, \aap, 429,
  887

\bibitem[{{Percival} {et~al.}(2003){Percival}, {Salaris}, \&
  {Kilkenny}}]{per03a}
{Percival}, S.~M., {Salaris}, M., \& {Kilkenny}, D. 2003, \aap, 400, 541

\bibitem[{{Perryman} {et~al.}(1998){Perryman}, {Brown}, {Lebreton}, {Gomez},
  {Turon}, {de Strobel}, {Mermilliod}, {Robichon}, {Kovalevsky}, \&
  {Crifo}}]{perr98}
{Perryman}, M.~A.~C., {Brown}, A.~G.~A., {Lebreton}, Y., {Gomez}, A., {Turon},
  C., {de Strobel}, G.~C., {Mermilliod}, J.~C., {Robichon}, N., {Kovalevsky},
  J., \& {Crifo}, F. 1998, \aap, 331, 81

\bibitem[{{Pietrinferni} {et~al.}(2004){Pietrinferni}, {Cassisi}, {Salaris}, \&
  {Castelli}}]{pietr04}
{Pietrinferni}, A., {Cassisi}, S., {Salaris}, M., \& {Castelli}, F. 2004, \apj,
  612, 168

\bibitem[{{Prada Moroni} \& {Straniero}(2007)}]{prad07}
{Prada Moroni}, P.~G. \& {Straniero}, O. 2007, \aap, 466, 1043

\bibitem[{{Press} {et~al.}(1992){Press}, {Teukolsky}, {Vetterling}, \&
  {Flannery}}]{ptvf92}
{Press}, W.~H., {Teukolsky}, S.~A., {Vetterling}, W.~T., \& {Flannery}, B.~P.
  1992, {Numerical recipes in FORTRAN. The art of scientific computing}
  (Cambridge: University Press, |c1992, 2nd ed.)

\bibitem[{{Provencal} {et~al.}(1998){Provencal}, {Shipman}, {Hog}, \&
  {Thejll}}]{prov98}
{Provencal}, J.~L., {Shipman}, H.~L., {Hog}, E., \& {Thejll}, P. 1998, \apj,
  494, 759

\bibitem[{{Raffelt}(1996)}]{raf_book}
{Raffelt}, G.~G. 1996, {Stars as laboratories for fundamental physics: the
  astrophysics of neutrinos, axions, and other weakly interacting particles}
  (Chicago, University of Chicago Press)

\bibitem[{{Reid}(1996)}]{reid96}
{Reid}, I.~N. 1996, \aj, 111, 2000

\bibitem[{{Rubin} {et~al.}(2008){Rubin}, {Williams}, {Bolte}, \&
  {Koester}}]{rubin08}
{Rubin}, K.~H.~R., {Williams}, K.~A., {Bolte}, M., \& {Koester}, D. 2008, \aj,
  135, 2163

\bibitem[{{Salaris} {et~al.}(2001){Salaris}, {Cassisi}, {Garc{\'{\i}}a-Berro},
  {Isern}, \& {Torres}}]{scg01}
{Salaris}, M., {Cassisi}, S., {Garc{\'{\i}}a-Berro}, E., {Isern}, J., \&
  {Torres}, S. 2001, \aap, 371, 921

\bibitem[{{Salaris} {et~al.}(1997){Salaris}, {Dominguez}, {Garcia-Berro},
  {Hernanz}, {Isern}, \& {Mochkovitch}}]{sal97}
{Salaris}, M., {Dominguez}, I., {Garcia-Berro}, E., {Hernanz}, M., {Isern}, J.,
  \& {Mochkovitch}, R. 1997, \apj, 486, 413

\bibitem[{{Salaris} {et~al.}(2000){Salaris}, {Garc{\'{\i}}a-Berro}, {Hernanz},
  {Isern}, \& {Saumon}}]{sala00}
{Salaris}, M., {Garc{\'{\i}}a-Berro}, E., {Hernanz}, M., {Isern}, J., \&
  {Saumon}, D. 2000, \apj, 544, 1036

\bibitem[{{Schaller} {et~al.}(1992){Schaller}, {Schaerer}, {Meynet}, \&
  {Maeder}}]{schall92}
{Schaller}, G., {Schaerer}, D., {Meynet}, G., \& {Maeder}, A. 1992, \aaps, 96,
  269

\bibitem[{{Schuler} {et~al.}(2003){Schuler}, {King}, {Fischer}, {Soderblom}, \&
  {Jones}}]{schul03}
{Schuler}, S.~C., {King}, J.~R., {Fischer}, D.~A., {Soderblom}, D.~R., \&
  {Jones}, B.~F. 2003, \aj, 125, 2085

\bibitem[{{Serenelli} \& {Fukugita}(2007)}]{Serfuk07}
{Serenelli}, A.~M. \& {Fukugita}, M. 2007, \apjs, 172, 649

\bibitem[{{Straniero} {et~al.}(2003){Straniero}, {Dom{\'{\i}}nguez},
  {Imbriani}, \& {Piersanti}}]{stra03}
{Straniero}, O., {Dom{\'{\i}}nguez}, I., {Imbriani}, G., \& {Piersanti}, L.
  2003, \apj, 583, 878

\bibitem[{{Sung} \& {Bessell}(1999)}]{sung99}
{Sung}, H. \& {Bessell}, M.~S. 1999, \mnras, 306, 361

\bibitem[{{Sung} {et~al.}(2002){Sung}, {Bessell}, {Lee}, \& {Lee}}]{sung02}
{Sung}, H., {Bessell}, M.~S., {Lee}, B.-W., \& {Lee}, S.-G. 2002, \aj, 123, 290

\bibitem[{{von Hippel}(2005)}]{hippel05}
{von Hippel}, T. 2005, \apj, 622, 565

\bibitem[{{Wagenhuber} \& {Groenewegen}(1998)}]{wg98}
{Wagenhuber}, J. \& {Groenewegen}, M.~A.~T. 1998, \aap, 340, 183

\bibitem[{{Weidemann}(1977)}]{weid77}
{Weidemann}, V. 1977, \aap, 59, 411

\bibitem[{{Weidemann}(1987)}]{weid87}
---. 1987, \aap, 188, 74

\bibitem[{{Weidemann}(2000)}]{weide00}
---. 2000, \aap, 363, 647

\bibitem[{{Weidemann} \& {Koester}(1983)}]{weidkoe83}
{Weidemann}, V. \& {Koester}, D. 1983, \aap, 121, 77

\bibitem[{{Williams} {et~al.}(2004){Williams}, {Bolte}, \& {Koester}}]{wbk04}
{Williams}, K.~A., {Bolte}, M., \& {Koester}, D. 2004, \apjl, 615, L49

\end{thebibliography}

\clearpage

\begin{deluxetable}{llccccl}
\tablewidth{0pt}
\tablecaption{Adopted data for our WD sample, taken from the
  literature (see Sect.~2.1).\label{WDsample}}
\tablehead{
\colhead{System}  & \colhead{WD name}      & \colhead{\teff\ [K]}      &
\colhead{$\sigma(T_{\rm eff})$} & \colhead{$\log(g)$ [${\rm cm \ s^{-2}}$]}  &
\colhead{$\sigma(\log(g))$} & \colhead{ID}}
\startdata
\tableline\tableline
 Pleiades  & LB1497 &32841   &   170   &     8.630   &   0.040 & 1\\
\tableline
  Hyades   & 0352+098 &16630   &   350    &    8.160    &  0.050 & 2\\
 & 0406+169& 15180   &   350    &    8.300    &  0.050 & 3\\
 & 0421+162& 19570   &   350    &    8.090    &  0.050 & 4\\
 & 0425+168& 24420   &   350    &    8.110    &  0.050 & 5\\
 & 0431+125& 21340   &   350    &    8.040    &  0.050 & 6\\
 & 0438+108& 27390   &   350    &    8.070    &  0.050 & 7\\
 & 0437+138& 15340   &   350    &    8.260    &  0.050 & 8\\
\tableline
 Praesepe  & 0836+197&21950   &   350    &    8.450    &  0.050 & 9\\
 & 0836+201& 16630   &   350    &    8.010    &  0.050 & 10\\
 & 0836+199& 14060   &   630    &    8.340    &  0.060 & 11\\
 & 0837+199& 17100   &   350    &    8.320    &  0.050 & 12\\
 & 0837+218& 16833   &   250    &    8.390    &  0.030 & 13\\
 & 0837+185& 14748   &   400    &    8.240    &  0.050 & 14\\
 & 0840+200& 14180   &   350    &    8.230    &  0.050 & 15\\
 & 0833+194& 14999   &   250    &    8.180    &  0.040 & 16\\
 & 0840+190& 14765   &   270    &    8.210    &  0.030 & 17\\
 & 0840+205& 14527   &   390    &    8.240    &  0.040 & 18\\
 & 0843+184& 14498   &   200    &    8.220    &  0.040 & 19\\
\tableline
 NGC2516  & 2516-1 & 28170   &   310    &    8.480    &  0.170 & 20\\
 & 2516-2& 34200   &   610    &    8.600    &  0.110 & 21\\
 & 2516-3& 26870   &   330    &    8.550    &  0.070 & 22\\
 & 2516-5& 30760   &   420    &    8.700    &  0.120 & 23\\
\tableline
 NGC3532  &  3532-8& 23370   &  1065     &   7.713    &  0.148 & 24\\
 & 3532-9& 29800   &   616     &   7.827    &  0.229 & 25\\
 & 3532-10& 19270   &   974     &   8.143    &  0.266 & 26\\
\tableline
 NGC2099 (M37) & 2099-WD2& 19900   &   900     &   8.110    &  0.160 & 27\\
 & 2099-WD3& 18300   &   900     &   8.230    &  0.210 & 28\\
 & 2099-WD4& 16900   &  1100     &   8.400    &  0.260 & 29\\
 & 2099-WD5& 18300   &  1000     &   8.330    &  0.220 & 30\\
 & 2099-WD7& 17800   &  1400     &   8.420    &  0.320 & 31\\
 & 2099-WD9& 15300   &   400     &   8.000    &  0.080 & 32\\
 & 2099-WD10& 19300   &   400     &   8.200    &  0.070 & 33\\
 & 2099-WD11& 23000   &   600     &   8.540    &  0.100 & 34\\
 & 2099-WD12& 13300   &  1000     &   7.910    &  0.120 & 35\\
 & 2099-WD13& 18200   &   400     &   8.270    &  0.080 & 36\\
 & 2099-WD14& 11400   &   200     &   7.730    &  0.160 & 37\\
 & 2099-WD16& 13100   &   500     &   8.340    &  0.100 & 38\\
\tableline
NGC2168 (M35) & NGC2168 LAWDS1& 32400   &   512     &   8.400    &  0.125 & 39\\
  & NGC2168 LAWDS 2& 32700   &   603     &   8.340    &  0.080 & 40\\
  & NGC2168 LAWDS 5&52600   &  1160     &   8.240    &  0.095 & 41\\
  & NGC2168 LAWDS 6&55200   &   897     &   8.280    &  0.065 & 42\\
  & NGC2168 LAWDS 15&29900   &   318     &   8.480    &  0.060 & 43\\
  & NGC2168 LAWDS 27&30500   &   397     &   8.520    &  0.061 & 44\\
\tableline
Sirius   & Sirius B & 25193    &   37     &   8.566    &  0.010 & 45\\
\tableline
NGC7789 & NGC 7789-5 & 31213    &  238     &   7.904    &  0.054 & 46\\
        & NGC 7789-8 & 24319    &  447     &   8.004    &  0.066 & 47\\
        & NGC 7789-9 & 20939    &  727     &   7.838    &  0.115 & 48\\
\tableline
NGC6819 &  NGC 6819-6& 21094    &  252     &   7.832    &  0.036 & 49\\
        &  NGC 6819-7& 15971    &  197     &   7.908    &  0.038 & 50\\
\tableline
NGC 1039 & NGC1039 LAWDS 15& 25900 & 1100 & 8.380 & 0.120 & 51 \\
         & NGC1039 LAWDS 17& 24700 & 1100 & 8.440 & 0.120 & 52 \\
         & NGC1039 LAWDS S2& 31200 & 1100 & 8.320 & 0.120 & 53 \\
\tableline
\enddata
\end{deluxetable}

\clearpage

\begin{deluxetable}{llll}
\tablewidth{0pt}
\tablecaption{Sources of the data used for determining the
  cluster ages.\label{clsource}}
\tablehead{
\colhead{Name}               & \colhead{[Fe/H]}      &
\colhead{E(B-V)}          & \colhead{CMD}}
\startdata
\tableline\tableline
 Pleiades     & \citet{grat00} & \citet{per03a}& \citet{j58}\\
 Hyades       & \citet{grat00} &               & \citet{j55}\\
 Praesepe     & \citet{grat00} &               & \citet{j52}\\
 NGC2516      & \citet{grat00} & \citet{lot01} & \citet{sung02}\\
 NGC3532      & \citet{grat00} & \citet{lot01} & \citet{fern80}\\
 NGC2099 (M37)& \citet{chen03} & \citet{lot01} & \citet{sag02}\\
 NGC2168 (M35)& \citet{grat00} & \citet{lot01} & \citet{sung99}\\
 NGC7789      & \citet{grat00} & \citet{per03b}& \citet{mk99}\\
 NGC6819      & \citet{brag01} & \citet{brag01}& \citet{kalir01}\\
 NGC1039      & \citet{schul03}& \citet{lot01} & \citet{jonpros96} \\
\tableline
\enddata
\end{deluxetable}

\clearpage

\begin{deluxetable}{lrlrlll}
\tablewidth{0pt}
\tablecaption{Cluster ages determined from the adopted metallicities, reddening values and derived distance moduli, 
  using three different
  sets of isochrones.\label{Ages}}
\tablehead{
\colhead{Name}               & \colhead{[Fe/H]}      &
\colhead{\ebv} &  \colhead{$(m-M)_V$} &
\colhead{Age[Myr]\tablenotemark{a}}  &
\colhead{Age[Myr]\tablenotemark{b}} &
\colhead{Age[Myr]\tablenotemark{c}}}
\startdata
\tableline\tableline
 Pleiades & $-0.03\pm0.06$& 0.04               & 5.74$\pm$0.05     &
 50$\pm$10 & 85$\pm$10 & 95$\pm$10\\  
 Hyades & $ 0.13\pm0.06$& 0.0                  & 3.33$\pm$0.01     &
 440$\pm$40 & 640$\pm$40 & 630$\pm$40\\ 
 Praesepe & $ 0.04\pm0.06$& 0.0                & 6.24$\pm$0.04     &
 450$\pm$40 & 650$\pm$50 & 640$\pm$40\\ 
 NGC2516 & $-0.16\pm0.11$& 0.10                & 8.27$\pm$0.07     &
 85$\pm$45 & 130$\pm$50 & 140$\pm$50\\ 
 NGC3532 & $ 0.02\pm0.06$& 0.04                & 8.40$\pm$0.25     &
 300$\pm$100& 400$\pm$100& 400$\pm$100\\ 
 NGC2099 (M37) \tablenotemark{d}& $ 0.09\pm0.15$& 0.30
 &12.00$\pm$0.12     & 220$\pm$30 & 320$\pm$30 & 320$\pm$30\\ 
               & $-$0.20       & 0.23          &11.40$\pm$0.12     &
 350$\pm$40 & 550$\pm$50 & 540$\pm$50\\ 
 NGC2168 (M35) & $-0.19\pm0.15$& 0.26          &10.50$\pm$0.12     &
 85$\pm$25 & 120$\pm$30 & 130$\pm$30\\ 
 NGC7789 & $-0.13\pm0.08$& 0.29                &12.12$\pm$0.12     &
 1100$\pm$100&1500$\pm$100 &1600$\pm$100\\ 
 NGC6819   &    $   0.09\pm0.03$& $0.14\pm0.04$&12.60$\pm$0.20     &
 1500$\pm$200&2000$\pm$200  &2000$\pm$200\\ 
 NGC1039   & $0.07\pm0.04$ & 0.07              & 8.80$\pm$0.15      &
 150$\pm30$ & 250$\pm25$ & 250$\pm25$ \\ 
\tableline
\tableline
Sirius~A & $0.0$ & ----- & 181$\pm$30 & 170$\pm$50 & 170$\pm$25 \\
\tableline
\enddata
\tablenotetext{a}{Ages from BaSTI isochrones without overshooting}
\tablenotetext{b}{Ages from BaSTI isochrones with overshooting}
\tablenotetext{c}{Ages from Padua isochrones with overshooting}
\tablenotetext{d}{For NGC2099 a second alternative [Fe/H] value was
  tested (see text for details)} 
\end{deluxetable}

\clearpage

\begin{deluxetable}{clccccclcccccc}
\tablewidth{0pt}
\tablecaption{Fractional $1-\sigma$  uncertainties in  WD mass and  cooling age
  for relevant  WD input  physics, as functions  of effective  temperature and
  surface gravity. \label{tab:uncert}} 
\tablehead{
\colhead{} & 
\colhead{} & 
\multicolumn{5}{c}{Fract. uncert. for WD mass} &
\colhead{}  &
\multicolumn{5}{c}{Fract. uncert. for WD cooling age} \\
\cline{1-1} \cline{3-7} \cline{9-13}
\colhead{$\log(g)$~[${\rm cm \ s^{-2}}$] = } &
\colhead{} &
\colhead{8.6} & 
\colhead{8.4} &
\colhead{8.2} &
\colhead{8.0} &
\colhead{7.8} &
\colhead{} &
\colhead{8.6} & 
\colhead{8.4} &
\colhead{8.2} &
\colhead{8.0} &
\colhead{7.8} \\
\cline{1-1}
\colhead{$\log(T_{\rm eff})$~[K]}
}
\startdata
 & & \multicolumn{11}{c}{Cooling tracks} \\
4.60 &  &  0.017 & 0.015 & 0.016 & 0.016 & 0.003 & & 0.826 & 0.564 & 0.287
& 0.335 & 0.377 \\ 
4.45 & & 0.016 & 0.015 & 0.014 & 0.010 & 0.011 & & 0.036 & 0.262 & 0.291
& 0.174 & 0.006 \\ 
4.30 & & 0.015 & 0.014 & 0.013 & 0.007 & 0.017 & & 0.072 & 0.064 & 0.012
& 0.057 & 0.159 \\ 
4.15 & & 0.012 & 0.012 & 0.011 & 0.003 & 0.037 & & 0.035 & 0.114 & 0.080
& 0.029 & 0.100 \\ 
4.00 & & 0.009 & 0.006 & 0.010 & 0.000 & 0.061 & & 0.368 & 0.020 & 0.073
& 0.064 & 0.122 \\ 
\tableline
& & \multicolumn{11}{c}{Neutrino cooling} \\
4.60 & & 0.000 & 0.001 & 0.001 & 0.002 & 0.003 & & 0.700 & 0.396 & 0.320
& 0.269 & 0.217 \\
4.45 & & 0.000 & 0.000 & 0.000 & 0.001 & 0.001 & & 0.147 & 0.349 & 0.474
& 0.374 & 0.305 \\
4.30 & & 0.000 & 0.000 & 0.000 & 0.000 & 0.000 & & 0.042 & 0.075 & 0.137
& 0.250 & 0.326 \\
4.15 & & 0.000 & 0.000 & 0.000 & 0.000 & 0.000 & & 0.017 & 0.027 & 0.044
& 0.072 & 0.098 \\
4.00 & & 0.000 & 0.000 & 0.000 & 0.000 & 0.000 & & 0.007 & 0.011 & 0.018
& 0.027 & 0.037 \\
\tableline
& & \multicolumn{11}{c}{Conductive opacity} \\
4.60 & & 0.000 & 0.000 & 0.000 &
0.001 & 0.001 & & 0.391 & 0.095 & 0.040 & 0.022 & 0.020 \\
4.45 & & 0.000 & 0.000 & 0.001 & 0.001 & 0.002 & & 0.056 &
0.197 & 0.200 & 0.089 & 0.046 \\ 
4.30 & & 0.000 & 0.000 & 0.000 &
0.001 & 0.002 & & 0.033 & 0.009 & 0.028 & 0.083 & 0.090 \\
4.15 & & 0.000 & 0.000 & 0.000 & 0.000
 & 0.001 & & 0.056 & 0.049 & 0.040 & 0.024 & 0.035 \\
4.00 & & 0.000 & 0.000 & 0.000 & 0.000 & 0.000 & & 0.067 & 0.058 & 0.054 &
0.053 & 0.083 \\
\tableline
& & \multicolumn{11}{c}{Core composition} \\
4.60 & & 0.002 & 0.002 & 0.002 & 0.002 & 0.002 & & 0.058 & 0.046 & 0.043
& 0.046 & 0.066 \\
4.45 & & 0.002 & 0.002 & 0.002 & 0.002 & 0.002 & & 0.013 & 0.042 & 0.040
& 0.043 & 0.047 \\
4.30 & & 0.002 & 0.001 & 0.001 & 0.002 & 0.002 & & 0.003 & 0.028 & 0.036
& 0.040 & 0.043 \\
4.15 & & 0.002 & 0.001 & 0.001 & 0.001 & 0.002 & & 0.005 & 0.013 & 0.033
& 0.037 & 0.039 \\
4.00 & & 0.002 & 0.001 & 0.001 & 0.001 & 0.002 & & 0.001 & 0.007 & 0.029
& 0.037 & 0.039 \\
\tableline
& & \multicolumn{11}{c}{H-envelope thickness} \\
4.60 & & 0.006 & 0.008 & 0.012 & 0.014 & 0.020 & & 0.035 & 0.003 & 0.011
& 0.001 & 0.024 \\
4.45 & & 0.005 & 0.007 & 0.010 & 0.013 & 0.021 & & 0.029 & 0.015 & 0.007
& 0.004 & 0.014 \\
4.30 & & 0.005 & 0.006 & 0.009 & 0.012 & 0.024 & & 0.033 & 0.011 & 0.005
& 0.013 & 0.006 \\
4.15 & & 0.004 & 0.006 & 0.008 & 0.011 & 0.027 & & 0.028 & 0.027 & 0.007
& 0.008 & 0.015 \\
4.00 & & 0.003 & 0.005 & 0.007 & 0.010 & 0.028 & & 0.070 & 0.057 & 0.044
& 0.025 & 0.016 \\
\tableline
\enddata
\tablenotetext{a}{Columns and rows are labeled according to $\log{(g)}$
  and $\log{(T_{\rm eff})}$ respectively. Definitions of how the
  uncertainties were computed 
  are  given   in  Eqs.~\ref{eq:wdsys}-\ref{eq:wdenv}.  Note,   however,  that
  for the sake of readability, uncertainties given here are linear (not
  logarithmic) in mass and
  cooling age and have been assumed symmetric. Fractional
  uncertainties  smaller   than  0.001  are  shown  as   null  values.} 
\end{deluxetable}

\clearpage

\begin{deluxetable}{lccccccc}
\tablewidth{0pt}
\tablecaption{White dwarf cooling ages and associated mass (in solar
  mass units) determined 
  with the S00 cooling tracks. The bolometric  luminosity is obtained
  using the derived cluster distances. \label{WDage1}}
\tablehead{
\colhead{ID}&
\colhead{$\log{t_{\rm cool}}$}[yr]&
\colhead{$\sigma_{-}(\log{t_{\rm cool}})$} &
\colhead{$\sigma_{+}(\log{t_{\rm cool}})$} &
\colhead{$M_{\rm f}  [M_{\odot}]$} &
\colhead{$\sigma_{-}{(M_{\rm f})}$} &
\colhead{$\sigma_{+}{(M_{\rm f})}$} &
\colhead{$\log(L/L_{\odot})$} 
}
\startdata
\tableline\tableline
\multicolumn{2}{l}{Pleiades} \\
1 & 7.667 & $-$0.185 & 0.162 & 1.028 & $-$0.031 & 0.031 & $-$1.162 \\
\tableline
\multicolumn{2}{l}{Hyades} \\
2 & 8.262 & $-$0.063 & 0.056 & 0.713 & $-$0.031 & 0.031 & $-$2.032 \\
3 & 8.466 & $-$0.066 & 0.064 & 0.798 & $-$0.032 & 0.033 & $-$2.282 \\
4 & 7.981 & $-$0.105 & 0.086 & 0.679 & $-$0.031 & 0.031 & $-$1.701 \\
5 & 7.585 & $-$0.185 & 0.174 & 0.699 & $-$0.031 & 0.031 & $-$1.324 \\
6 & 7.783 & $-$0.146 & 0.130 & 0.652 & $-$0.032 & 0.032 & $-$1.518 \\
7 & 7.274 & $-$0.186 & 0.184 & 0.684 & $-$0.032 & 0.031 & $-$1.094 \\
8 & 8.425 & $-$0.063 & 0.061 & 0.773 & $-$0.032 & 0.033 & $-$2.237 \\
\tableline
\multicolumn{2}{l}{Praesepe} \\
9 & 8.113 & $-$0.072 & 0.065 & 0.901 & $-$0.034 & 0.034 & $-$1.739 \\
10& 8.159 & $-$0.076 & 0.066 & 0.622 & $-$0.028 & 0.032 & $-$1.942 \\
11& 8.584 & $-$0.086 & 0.085 & 0.820 & $-$0.040 & 0.039 & $-$2.443 \\
12& 8.338 & $-$0.064 & 0.062 & 0.812 & $-$0.034 & 0.034 & $-$2.087 \\
13& 8.408 & $-$0.058 & 0.057 & 0.857 & $-$0.023 & 0.022 & $-$2.161 \\
14& 8.459 & $-$0.064 & 0.063 & 0.759 & $-$0.032 & 0.033 & $-$2.294 \\
15& 8.498 & $-$0.062 & 0.062 & 0.751 & $-$0.033 & 0.033 & $-$2.357 \\
16& 8.402 & $-$0.053 & 0.051 & 0.722 & $-$0.026 & 0.027 & $-$2.226 \\
17& 8.439 & $-$0.052 & 0.051 & 0.740 & $-$0.021 & 0.021 & $-$2.273 \\
18& 8.476 & $-$0.061 & 0.061 & 0.759 & $-$0.027 & 0.027 & $-$2.320 \\
19& 8.467 & $-$0.054 & 0.053 & 0.746 & $-$0.027 & 0.027 & $-$2.311 \\
\tableline
\multicolumn{2}{l}{NGC2516} \\
20& 7.746 & $-$0.298 & 0.192 & 0.926 & $-$0.106 & 0.105 & $-$1.323 \\
21& 7.543 & $-$0.364 & 0.297 & 1.008 & $-$0.071 & 0.070 & $-$1.069 \\
22& 7.888 & $-$0.116 & 0.087 & 0.970 & $-$0.045 & 0.045 & $-$1.455 \\
23& 7.858 & $-$0.202 & 0.154 & 1.075 & $-$0.078 & 0.078 & $-$1.326 \\
\tableline
\multicolumn{2}{l}{NGC3532} \\
24& 7.373 & $-$0.216 & 0.219 & 0.482 & $-$0.085 & 0.076 & $-$1.164 \\
25& 7.064 & $-$0.303 & 0.161 & 0.563 & $-$0.169 & 0.145 & $-$0.789 \\
26& 8.047 & $-$0.284 & 0.213 & 0.708 & $-$0.150 & 0.164 & $-$1.762 \\
\tableline
\multicolumn{2}{l}{NGC2099 (M37)} \\
27 & 7.971 & $-$0.215 & 0.150 & 0.690 & $-$0.093 & 0.096 & $-$1.685 \\
28 & 8.185 & $-$0.193 & 0.167 & 0.759 & $-$0.128 & 0.133 & $-$1.909 \\
29 & 8.411 & $-$0.214 & 0.245 & 0.864 & $-$0.163 & 0.164 & $-$2.161 \\
30 & 8.260 & $-$0.190 & 0.194 & 0.821 & $-$0.135 & 0.139 & $-$1.975 \\
31 & 8.366 & $-$0.268 & 0.305 & 0.877 & $-$0.198 & 0.203 & $-$2.085 \\
32 & 8.264 & $-$0.083 & 0.072 & 0.614 & $-$0.044 & 0.051 & $-$2.082 \\
33 & 8.090 & $-$0.094 & 0.074 & 0.742 & $-$0.044 & 0.044 & $-$1.796 \\
34 & 8.111 & $-$0.108 & 0.097 & 0.959 & $-$0.065 & 0.064 & $-$1.720 \\
35 & 8.380 & $-$0.133 & 0.120 & 0.561 & $-$0.080 & 0.067 & $-$2.275 \\
36 & 8.222 & $-$0.082 & 0.075 & 0.783 & $-$0.050 & 0.052 & $-$1.945 \\
37 & 8.455 & $-$0.125 & 0.121 & 0.448 & $-$0.129 & 0.123 & $-$2.461 \\
38 & 8.667 & $-$0.097 & 0.103 & 0.818 & $-$0.064 & 0.066 & $-$2.567 \\
\tableline
\multicolumn{2}{l}{NGC2168 (M35)} \\
39 & 7.343 & $-$0.339 & 0.355 & 0.884 & $-$0.078 & 0.079 & $-$1.021 \\
40 & 7.223 & $-$0.310 & 0.327 & 0.849 & $-$0.051 & 0.050 & $-$0.962 \\
41 & 6.351 & $-$0.213 & 0.208 & 0.820 & $-$0.055 & 0.056 & $-$0.052 \\
42 & 6.292 & $-$0.218 & 0.217 & 0.845 & $-$0.041 & 0.040 & $+$0.006 \\
43 & 7.634 & $-$0.219 & 0.200 & 0.928 & $-$0.040 & 0.040 & $-$1.219 \\
44 & 7.644 & $-$0.215 & 0.186 & 0.955 & $-$0.041 & 0.041 & $-$1.212 \\
\tableline
\multicolumn{2}{l}{Sirius B} \\
45 & 7.996 & $-$0.063 & 0.062 & 0.978 & $-$0.016 & 0.016 & $-$1.580 \\
\tableline
\multicolumn{2}{l}{NGC7789} \\
46 & 6.998 & $-$0.137 & 0.135 & 0.601 & $-$0.028 & 0.029 & $-$0.757 \\
47 & 7.480 & $-$0.181 & 0.179 & 0.637 & $-$0.037 & 0.039 & $-$1.265 \\
48 & 7.627 & $-$0.205 & 0.192 & 0.543 & $-$0.059 & 0.058 & $-$1.428 \\
\tableline
\multicolumn{2}{l}{NGC6819} \\
49 & 7.609 & $-$0.153 & 0.150 & 0.540 & $-$0.021 & 0.019 & $-$1.412 \\
50 & 8.141 & $-$0.071 & 0.063 & 0.566 & $-$0.023 & 0.021 & $-$1.951 \\
\tableline
\multicolumn{2}{l}{NGC1039} \\
51 & 7.801 & $-$0.239 & 0.161 & 0.863 & $-$0.075 & 0.075 & $-$1.443 \\
52 & 7.934 & $-$0.179 & 0.126 & 0.899 & $-$0.076 & 0.076 & $-$1.533 \\
53 & 7.308 & $-$0.312 & 0.339 & 0.834 & $-$0.072 & 0.076 & $-$1.144 \\
\tableline

\enddata
\end{deluxetable}

\begin{deluxetable}{ccccccccccc}
\tablewidth{0pt}
\tablecaption{Initial--final--mass-relation obtained by combining BaSTI
  progenitor models with and without
overshooting (indicated by OV resp.\ non-OV) with S00 WD tracks 
\tablenotemark{a}.
\label{MiMftab}}
\tablehead{
\colhead{} & \colhead{} & \colhead{} & \colhead{} & \multicolumn{3}{c}{OV models} & \colhead{} &
\multicolumn{3}{c}{non-OV models}  \\
\cline{5-7}  \cline{9-11}
\colhead{ID} &
\colhead{$M_{\rm f}  [M_{\odot}]$} &
\colhead{$\sigma(M_{\rm f})$} &
\colhead{} &
\colhead{$M_{\rm i} [M_{\odot}]$} &
\colhead{$\sigma_-(M_{\rm i})$} &
\colhead{$\sigma_+(M_{\rm i})$} &
\colhead{} &
\colhead{$M_{\rm i} [M_{\odot}]$} &
\colhead{$\sigma_-(M_{\rm i})$} &
\colhead{$\sigma_+(M_{\rm i})$} 
 }
\startdata
\tableline\tableline
\multicolumn{11}{l}{Pleiades} \\
1 & 1.028 & 0.031 && 7.847 & $-$1.257 & 3.069 && 15.52 & $-$3.986 & 99.90 \\
\tableline
\multicolumn{11}{l}{Hyades} \\
2 & 0.713 & 0.031 && 3.067 & $-$0.100 & 0.107 && 3.690 & $-$0.192 & 0.224  \\
3 & 0.798 & 0.032 && 3.383 & $-$0.174 & 0.208 && 4.438 & $-$0.454 & 0.813  \\
4 & 0.679 & 0.031 && 2.887 & $-$0.076 & 0.079 && 3.385 & $-$0.128 & 0.142  \\
5 & 0.699 & 0.031 && 2.771 & $-$0.067 & 0.066 && 3.184 & $-$0.108 & 0.111  \\
6 & 0.652 & 0.032 && 2.815 & $-$0.071 & 0.071 && 3.261 & $-$0.116 & 0.122  \\
7 & 0.684 & 0.031 && 2.746 & $-$0.063 & 0.064 && 3.144 & $-$0.097 & 0.105  \\
8 & 0.773 & 0.032 && 3.284 & $-$0.151 & 0.172 && 4.240 & $-$0.349 & 0.549  \\
\tableline
\multicolumn{11}{l}{Praesepe} \\
9  & 0.901 & 0.034 && 2.799 & $-$0.094 & 0.106 && 3.292 & $-$0.138 & 0.161 \\
10 & 0.622 & 0.030 && 2.827 & $-$0.099 & 0.112 && 3.341 & $-$0.149 & 0.179 \\
11 & 0.820 & 0.039 && 3.543 & $-$0.337 & 0.548 && 5.953 & $-$1.223 & 8.634 \\
12 & 0.812 & 0.034 && 2.972 & $-$0.130 & 0.148 && 3.655 & $-$0.230 & 0.287 \\
13 & 0.857 & 0.023 && 3.058 & $-$0.153 & 0.174 && 3.926 & $-$0.290 & 0.421 \\
14 & 0.759 & 0.032 && 3.179 & $-$0.183 & 0.221 && 4.151 & $-$0.400 & 0.696 \\
15 & 0.751 & 0.034 && 3.282 & $-$0.206 & 0.260 && 4.457 & $-$0.524 & 1.048 \\
16 & 0.722 & 0.027 && 3.045 & $-$0.146 & 0.167 && 3.901 & $-$0.272 & 0.386 \\
17 & 0.740 & 0.021 && 3.130 & $-$0.160 & 0.188 && 4.060 & $-$0.328 & 0.502 \\
18 & 0.759 & 0.027 && 3.223 & $-$0.189 & 0.232 && 4.233 & $-$0.447 & 0.801 \\
19 & 0.746 & 0.027 && 3.200 & $-$0.175 & 0.210 && 4.190 & $-$0.397 & 0.654 \\
\tableline
\multicolumn{11}{l}{NGC2516} \\
20 & 0.926 & 0.106 && 5.856 & $-$1.133 & 3.679 && 8.672 & $-$2.774 & 99.90 \\
21 & 1.008 & 0.071 && 5.269 & $-$0.837 & 1.899 && 6.732 & $-$1.511 & 5.739 \\
22 & 0.970 & 0.045 && 6.766 & $-$1.541 & 7.174 && 13.99 & $-$5.870 & 99.90 \\
23 & 1.075 & 0.078 && 6.492 & $-$1.415 & 5.640 && 12.35 & $-$4.929 & 99.90 \\
\tableline
\multicolumn{11}{l}{NGC3532} \\
24 & 0.482 & 0.081 && 3.132 & $-$0.252 & 0.369 && 3.450 & $-$0.340 & 0.563 \\
25 & 0.563 & 0.157 && 3.085 & $-$0.243 & 0.348 && 3.409 & $-$0.321 & 0.523 \\
26 & 0.708 & 0.157 && 3.466 & $-$0.399 & 0.620 && 3.975 & $-$0.560 & 1.316 \\
\tableline
\multicolumn{11}{l}{NGC2099 (M37); ${\rm [Fe/H]}=0.09$} \\
27 & 0.690 & 0.095 && 3.934 & $-$0.248 & 0.326 && 4.746 & $-$0.491 & 0.868 \\
28 & 0.759 & 0.131 && 4.372 & $-$0.436 & 0.940 && 6.009 & $-$1.110 & 6.012 \\
29 & 0.864 & 0.164 && 6.307 & $-$1.557 & 99.90 && 99.90 & 99.90    & 99.90 \\
30 & 0.821 & 0.137 && 4.666 & $-$0.567 & 1.891 && 7.628 & $-$2.020 & 99.90 \\
31 & 0.877 & 0.200 && 5.497 & $-$1.099 & 99.90 && 99.90 & 99.90    & 99.90 \\
32 & 0.614 & 0.048 && 4.689 & $-$0.437 & 0.693 && 7.806 & $-$1.834 & 8.624 \\
33 & 0.742 & 0.044 && 4.153 & $-$0.238 & 0.305 && 5.187 & $-$0.593 & 1.015 \\
34 & 0.959 & 0.065 && 4.198 & $-$0.274 & 0.382 && 5.323 & $-$0.691 & 1.433 \\
35 & 0.561 & 0.074 && 5.680 & $-$1.022 & 4.981 && 99.90 & 99.90    & 99.90 \\
36 & 0.783 & 0.051 && 4.471 & $-$0.364 & 0.519 && 6.574 & $-$1.206 & 3.976 \\
37 & 0.448 & 0.126 && 8.210 & $-$2.545 & 99.90 && 99.90 & 99.90    & 99.90 \\
38 & 0.818 & 0.065 && 99.90 &    99.90 & 99.90 && 99.90 & 99.90    & 99.90 \\
\tableline
\multicolumn{11}{l}{NGC2168 (M35)} \\
39 & 0.884 & 0.079 && 5.209 & $-$0.595 & 0.928 && 6.119 & $-$0.915 & 1.695 \\
40 & 0.849 & 0.050 && 5.111 & $-$0.522 & 0.796 && 5.936 & $-$0.784 & 1.330 \\
41 & 0.820 & 0.055 && 4.903 & $-$0.387 & 0.605 && 5.468 & $-$0.550 & 0.890 \\
42 & 0.845 & 0.040 && 4.899 & $-$0.385 & 0.602 && 5.461 & $-$0.549 & 0.885 \\
43 & 0.928 & 0.040 && 5.767 & $-$0.862 & 1.670 && 7.312 & $-$1.527 & 4.464 \\
44 & 0.955 & 0.041 && 5.804 & $-$0.869 & 1.669 && 7.383 & $-$1.558 & 4.441 \\
\tableline
\multicolumn{11}{l}{Sirius B} \\
45 & 0.978 & 0.016 && 5.934 & $-$1.139 & 3.904 && 5.426 & $-$0.777 & 1.198 \\
\tableline
\multicolumn{11}{l}{NGC7789} \\
46 & 0.601 & 0.029 && 1.853 & $-$0.096 & 0.128 && 1.954 & $-$0.072 & 0.105 \\
47 & 0.637 & 0.038 && 1.864 & $-$0.100 & 0.132 && 1.971 & $-$0.078 & 0.116 \\
48 & 0.543 & 0.059 && 1.871 & $-$0.101 & 0.134 && 1.983 & $-$0.083 & 0.125 \\
\tableline
\multicolumn{11}{l}{NGC6819} \\
49 & 0.540 & 0.020 && 1.736 & $-$0.090 & 0.117 && 1.852 & $-$0.073 & 0.077 \\
50 & 0.566 & 0.022 && 1.770 & $-$0.097 & 0.130 && 1.887 & $-$0.072 & 0.096 \\
\tableline
\multicolumn{11}{l}{NGC1039} \\
51 & 0.863 & 0.075 && 4.124 & $-$0.364 & 0.514 && 5.382 & $-$0.717 & 1.369 \\
52 & 0.899 & 0.076 && 4.319 & $-$0.435 & 0.659 && 6.069 & $-$1.006 & 2.652 \\
53 & 0.834 & 0.074 && 3.835 & $-$0.279 & 0.348 && 4.607 & $-$0.386 & 0.511 \\
\tableline
\tableline
\multicolumn{11}{l}{NGC2099 (M37); ${\rm [Fe/H]}=-0.20$} \\
27 & 0.690 & 0.095 && 2.994 & $-$0.133 & 0.145 && 3.643 & $-$0.269 & 0.349 \\
28 & 0.759 & 0.131 && 3.123 & $-$0.196 & 0.259 && 4.029 & $-$0.443 & 0.947 \\
29 & 0.864 & 0.163 && 3.522 & $-$0.355 & 1.056 && 5.248 & $-$1.460 & 99.90 \\
30 & 0.821 & 0.137 && 3.232 & $-$0.233 & 0.374 && 4.226 & $-$0.570 & 1.727 \\
31 & 0.877 & 0.201 && 3.424 & $-$0.325 & 1.085 && 4.814 & $-$1.064 & 99.90 \\
32 & 0.614 & 0.047 && 3.239 & $-$0.173 & 0.200 && 4.238 & $-$0.486 & 0.844 \\
33 & 0.742 & 0.043 && 3.055 & $-$0.128 & 0.144 && 3.823 & $-$0.273 & 0.382 \\
34 & 0.959 & 0.065 && 3.068 & $-$0.139 & 0.156 && 3.865 & $-$0.306 & 0.443 \\
35 & 0.561 & 0.074 && 3.454 & $-$0.282 & 0.412 && 4.923 & $-$1.010 & 4.628 \\
36 & 0.783 & 0.051 && 3.174 & $-$0.162 & 0.181 && 4.121 & $-$0.408 & 0.645 \\
37 & 0.448 & 0.125 && 3.628 & $-$0.361 & 0.628 && 6.066 & $-$2.242 & 99.90 \\
38 & 0.818 & 0.066 && 5.529 & $-$1.125 & 9.156 && 99.90 &    99.90 & 99.90 \\
\tableline
\enddata
\tablenotetext{a}{Values for initial masses  equal to $99.90\,M_{\odot}$ flag those cases
where the central value of the progenitor age $t_{\rm prog}$, is
negative.  Similarly, error values  of $99.90$  indicate that  progenitor ages
deviating 1-$\sigma$  from the central  value $t_{\rm prog}$ are  negative and
as   a    consequence   1-$\sigma$   values   for $M_{\rm i}$   cannot   be
obtained.}
\end{deluxetable}

\clearpage

\clearpage

\clearpage

\clearpage

\clearpage

\clearpage

\clearpage

\end{document}